\documentclass[prd,aps,showpacs,nofootinbib,nobibnotes]{revtex4}

\usepackage{graphicx}
\usepackage{epsfig,psfrag}

\def\gsim{\;\rlap{\lower 2.5pt 
\hbox{$\sim$}}\raise 1.5pt\hbox{$>$}\;}
\def\lsim{\;\rlap{\lower 2.5pt 
\hbox{$\sim$}}\raise 1.5pt\hbox{$<$}\;}

\begin {document}

\title{Dark Energy Constraints from Galaxy Cluster Peculiar Velocities}
\author{Suman Bhattacharya}
\email{sub5@pitt.edu}
\author{Arthur Kosowsky}
\email{kosowsky@pitt.edu}
\affiliation{Department of Physics and Astronomy, University of Pittsburgh, 3941 O'Hara Street, Pittsburgh, PA 15260 USA}

\begin{abstract}
Future multifrequency microwave background experiments with arcminute resolution and micro-Kelvin temperature sensitivity will be able to detect the kinetic Sunyaev-Zeldovich effect, providing a  way to measure radial peculiar velocities of massive galaxy clusters. We show that cluster peculiar velocities have the potential to constrain several dark energy parameters. We compare three velocity statistics (the distribution of radial velocities, the mean pairwise streaming velocity, and the velocity correlation function) and analyze the relative merits of these statistics in constraining dark energy parameters. 
Of the three statistics, mean pairwise streaming velocity provides constraints that are least sensitive to velocity errors: the constraints on parameters degrades only by a factor of two when the random error is increased from 100 to 500 km/s. We also compare cluster velocities with other dark energy probes proposed in the Dark Energy Task Force report. For cluster velocity measurements with realistic
priors, the eventual constraints on the dark energy density, the dark energy equation of state and its evolution are comparable to constraints from supernovae measurements, and better than cluster counts and baryon acoustic oscillations; 
adding velocity to other dark energy probes improves constraints on the figure of merit by more than a factor of two. For upcoming Sunyaev-Zeldovich galaxy cluster surveys,
even velocity measurements with errors as large as 1000 km/s will substantially improve
the cosmological constraints compared to using the cluster number density alone.
\end{abstract}

\maketitle

\section{Introduction}
\label{sec:intro}

Historically, the primary goal of cosmology has been determination of cosmological parameters
describing the overall properties of the universe. This quest has advanced greatly in the past decade
with the precise measurement of microwave background temperature fluctuations putting sharp 
constraints on many parameters \cite{jungman96,WMAP3}. Pushing cosmological parameter
determination to ever-greater precision might have been an academic pursuit, except for
the surprising discovery of the universe's accelerating expansion \cite{perlmutter98,riess98},
coupled with the discrepancy between the geometry of the universe \cite{geometry99,geometry00}
and its mass density 
\cite{shaya95,willick97,massden97,massden97_1,massden97_2,jus00}. 
The likely implications of  the inferred ``dark energy'' for fundamental physics are fueling a wide array of 
next-generation cosmological
experiments. Current measurements can constrain the dark energy density and its current equation of 
state at around the $10\%$ level from a combination of WMAP 3-year data \citep{WMAP3}
and the Supernova Legacy Survey 1-year data \citep{SNLS1}, for example. 
However, current experiments can only constrain dark energy density and its equation of state at the 
present epoch. We also need to quantify the evolution of dark energy with redshift, for this
determines whether the dark energy is fundamentally a dynamic (evolving scalar field) or
static (cosmological constant) entity. In recent years, a quartet of methods has emerged
as the most discussed for constraining dark energy redshift evolution: weak lensing by large-scale 
structure, primordial baryon acoustic oscillations (BAO) observed as a feature in the matter power 
spectrum at low redshifts, the distance-redshift relation measured via SNIa standard candles, and the 
redshift evolution of galaxy cluster counts detected via the Sunyaev-Zeldovich Effect. The relative
merits of these probes were considered in detail by the recent report from the Dark Energy
Task Force \citep{detf}. Each of these probes suffers from different sources of systematic error:  cluster 
counts are subject to uncertainty in the mass-SZ relation\citep{francis05}; weak lensing and BAO suffer 
from uncertainty in modelling baryon physics and nonlinear effects \citep{zentner07, zentner07_1, smith07}; and SNIa distance 
mesurements require knowledge about the extent to which the supernovas serve as
standardizable candles over a range of redshifts \citep{james06}. 
It is clearly important to probe 
cosmology through multiple techniques to check consistency between individual probes.

This paper addresses another approach to constraining dark energy which so far has received
comparatively little attention: the line-of-sight peculiar velocities of galaxy clusters. A moving galaxy 
cluster will induce a nearly-blackbody shift in the distribution of the microwave photons passing
through the cluster, due to Compton upscattering of the photons by hot electrons in the cluster gas
\citep{ksz80}. The temperature shift, known as the kinematic or kinetic Sunyaev-Zeldovich (kSZ) effect, is 
proportional to the line-of-sight momentum of the cluster gas (being linearly proportional
to both the optical depth for Compton scattering
and to the line-of-sight velocity of the cluster with respect to the microwave background rest frame),
while being independent of the cluster gas temperature. It is substantially smaller than
the more familiar thermal Sunyaev-Zeldovich effect \cite{sz80}; a typical large cluster
with a thermal SZ distortion of 100 $\mu$K at the frequency with the largest distortion may
have a kSZ signal of 5 to 10 $\mu$K for typical cluster velocities. The kSZ effect can be thought
of as essentially a Doppler shift due to cluster motion. Like the related thermal SZ effect,
the kSZ effect has the remarkable property that its imprint in the microwave background which we observe 
today is independent of the cluster's redshift, making it potentially an excellent probe
of cosmology. If we can reliably measure the kSZ effect in galaxy clusters, we expect the 
line-of-sight velocity error 
for an individual cluster to be largely independent of the cluster redshift, in marked contrast
to galaxy peculiar velocity surveys. 

The cosmological velocity field on cluster scales, arising solely from the effects of gravitational
instability in the universe, is a potent probe of structure formation. It is thus also potentially
a strong probe of dark energy, whose properties affect the rate of structure growth. Here
we study the feasibility of probing dark energy parameters using peculiar velocity of galaxy clusters 
obtainable through detection of kinematic Sunyaev-Zeldovich effect. The small amplitude of kSZ 
distortions, combined with the need to separate this small blackbody signal from several other larger 
signals with various spectra, makes detection challenging. Several studies have shown that in the 
absense of any foreground contamination, it would be possible to measure cluster velocities with  
reasonable accuracy ($\approx 100$ to 150 km/s) through multi-frequency SZ measurements with 
arcminute resolution, combined with X-ray followup \citep{sehgal05, diaferio05}. Even in the presence of 
point source contamination, it is possible to measure velocities with an accuracy of perhaps 200 km/s 
through multifrequency measurements with arcminute resolution \citep{knox04} and sufficient
integration time. Internal motions of the intracluster medium give an irreducible random error of around 100 km/s \citep{nagai03}. 

Upcoming SZ measurements like ACT \citep{kosowsky06, fowler07} and SPT  \citep{ruhl05} are designed 
to detect large numbers of clusters through their SZ signatures. The ACT collaboration foresees maps of 
sufficient raw sensitivity to measure the kSZ effect in many clusters, making detailed studies of the 
cosmological impact of future kSZ measurements timely.  Some recent work has shown that the 
kSZ correlation function will put significant constraints on the 
dark energy equation of state \citep{penn06}, 
and  cross-correlation of the kSZ signal with the galaxy density can constrain the redshift 
evolution of the equation of state \citep{dedeo06}. Cluster velocities alone can be used to constrain the 
matter density of the universe \citep{peel02, bk06}, the primordial power spectrum normalization \cite
{bk06}, and the dark energy equation of state \citep{bk06}. 

The goal of this paper is twofold. Following up on our initial study \citep{bk06}, 
we study the accuracy of theoretical models of cluster velocity statistics by comparing with numerical 
simulations, address error analysis in greater detail, and compare the relative merits of various velocity 
statistics in constraining dark energy parameters; we also compare cluster velocities with the Dark Energy
Task Force methods. We find that all three velocity statistics considered here can be computed using the 
halo model within likely measurement uncertainties. We then use a Fisher matrix calculation to compare
the power of various velocity statistics as dark energy probes over a range of velocity errors. Remarkably, for a sufficiently large velocity catalog, the dark 
energy parameter constraints degrade only by a factor of two when the velocity errors 
increase by a factor of five.  Comparing with other dark energy probes, cluster velocities from a large 
survey can provide dark energy constraints that are comparable to weak lensing and supernovae and a 
factor of two to three better than cluster counts and BAO. Combining cluster velocities with other dark 
energy probes improves the total constraint on the dark energy density by 10-15$\%$ and the Dark 
Energy Task Force Figure of Merit by a factor of 1.4 to 2.5. Cluster velocities can be competitive with other 
proposed techniques for probing dark energy, with  completely different systematic errors. 

Throughout this paper, we assume a cluster velocity catalog with some normal velocity
error; we consider errors from 200 km/sec to 1000 km/sec, representing
a range from optimistic to conservative based on current experimental sensitivities and
anticipated astrophysical complications. Using this range of errors, we then evaluate the statistical
constraints on dark energy parameters, assuming a cluster catalog with a given number of cluster
velocities. In practice, constraints from cluster velocities may well be dominated by systematic,
rather than statistical, errors, like all other methods of probing dark energy. When analyzing
real data to constrain dark energy, understanding these systematic errors is obviously crucial
to getting the right answers. For our purposes here, we only aim to evaluate the statistical
power of cluster velocities to constrain dark energy; we thus ignore systematic errors, keeping in
mind that any results here require an understanding of all relevant systematics to be realized
in practice. Note that systematic errors will tend to bias parameter constraints but will not
generally change the size of the statistical errors significantly. A discussion of various
relevant systematics is given in the last section of the paper; we will address this issue
in more detail elsewhere. 

This paper is organized as follows. Section II gives theoretical approximations of various velocity statistics 
computed using the halo model; Section III studies the accuracy of the halo model expressions by 
comparing with simulations. Section IV discusses various sources of errors for each of the statistics and 
presents analytic expressions for the errors; detailed derivations of these expressions 
are given in three Appendices. Using these expressions for the values of the velocity
statistics and their errors in hypothetical surveys of given sky area and velocity errors, 
Section V uses standard Fisher matrix techniques to compute constraints on dark energy parameters from 
the various velocity statistics. Section VI then compares the cosmological constraints obtainable
from  cluster velocities with those from the probes analyzed by the Dark Energy Task Force. Finally,  
Section VII discusses further refinement of the current calculations, including correlations
between various velocity statistics, extraction of cluster velocities from microwave
maps, and near-future prospects for kSZ velocity measurements. 
Throughout, we employ a standard spatially
flat $\Lambda$CDM model with parameters given by the best-fit WMAP 3-year values as
our fiducial cosmology unless otherwise noted.

\section{The Halo Model for Velocity Statistics}
\label{sec:statistics}

To study the potential of galaxy cluster velocity surveys to serve as a dark energy probe, we consider 
three different velocity statistics: the probability distribution function of the line-of-sight component of 
peculiar velocities $n_v$; the mean pairwise streaming velocity  $v_{ij}(r)$, which is the relative velocity 
along the line of separation of cluster pairs averaged over all pairs at fixed separation $r$; and the two-
point velocity correlation function $\langle v_iv_j\rangle(r)$ as a function of separation $r$. In the halo 
model picture of the dark matter distribution \cite{cooray02,zentnerref}, these quantities can be written as 
the sum of the contribution from one-halo and two-halo terms. However, we are interested only in very 
massive clusters ($M>10^{14} M_{\odot}$), so the one-halo term can be neglected. 

Here we summarize the halo model ingredients which go into computing the values of these 
velocity statistics for given cosmological models.
Define moments of the initial mass distribution with power spectrum $P(k)$ by \cite{bardeen86} 
\begin{equation}
\sigma_j^2(m)\equiv \frac{1}{2\pi^2}\int_0^{\infty}{dk ( k^{2+2j})P(k)W^2(kR(m))}
\label{eq:massmoments}
\end{equation}
when smoothed on the scale $R(m)= (3m/4\pi\rho_0)^{1/3}$ with the top-hat filter  
$W(x)=3[\sin(x)-x\cos(x)]/x^3$, and $\rho_0$ the present mean matter density. 
The spherical top-hat halo profile is adopted for simplicity. It could be replaced by a more realistic NFW 
profile; however, we are interested in statistics only of the most massive clusters at large scales where 
details of halo profiles make no significant difference. We also write
$H(a)$ for the Hubble parameter as a function of scale factor, $h\approx 0.7$ as the
Hubble parameter today in units of 100 km/s Mpc${}^{-1}$, 
and $R_{\rm local}$ for a smoothing scale with which the local background density $\delta$ is defined. 

The number density of halos of a given mass $n(m)$ is taken as the 
Jenkins mass function \cite{jenkins01}
\begin{equation}
\frac{dn}{dm}(m,z)= 0.315\frac{\rho_0}{m^2}\frac{d \ln \sigma_0(m)}{d \ln m}\exp\left[-\left|0.61-\ln(\sigma_0(m) D_a)\right|^{3.8}\right].
\end{equation}
This mass function is a fit to numerical simulations of cold dark matter gravitational clustering.
The bias factor can be written as \citep{sheth04} 
\begin{equation}
b(m,z)=1+\frac{\delta_{\rm crit}^2-\sigma_0^2(m)}{\sigma_0^2(m)\delta_{\rm crit}D_a}
\end{equation}
where $D_a$ is the linear growth factor at scale factor $a$, normalized to $1$ today, and the
critical overdensity $\delta_{\rm crit}\approx 1.686$. 
Since clusters preferentially form at points in space of larger overdensity, the number density of clusters 
for a given mass and formed in a given local overdensity can be written as \citep{sheth02}
\begin{equation}
n(m|\delta) \approx \left[1+b(m)\delta\right]{\bar n}(m).
\end{equation}  
The matter power spectrum $P(k)$ at the present epoch can be well fit through a transfer function as
\begin{equation}
P(k)= \frac{Bk^n}{\left[1+[\alpha k+(\beta k)^{3/2}+(\gamma k)^2]^{\nu}\right]^{2/\nu}}
\label{transferfunction}
\end{equation}
where $\alpha=(6.4/\Gamma) h^{-1}$ Mpc, $\beta=(3.0/\Gamma) h^{-1}$ Mpc, 
$\gamma=(1.7/\Gamma) h^{-1}$ Mpc, $\nu=1.13$ and $\Gamma=\Omega_m h$ \citep{bond84, bond92}. 
The normalization
$B$ is fixed at large scales by normalizing to the microwave background fluctuation amplitude.

\subsection{Probability Distribution Function}

The probability $p(v\,|\,m,\delta,a)$ that a cluster of mass $m$ located in an overdensity $\delta$ moves 
with a line-of-sight velocity $v$ can be approximated by a normal distribution \citep{sheth02},
\begin{equation} 
p(v\,|\,m,\delta,a)= \left(\frac{3}{2\pi}\right)^{1/2}\frac{1}{\sigma_v(m,a)}
\exp\left(-\frac{1}{2}\left[\frac{3v}{\sigma_v(m,a)}\right]^2\right)
\label{eq:pvmd}
\end{equation}
with  the three-dimensional velocity dispersion smoothed over a length scale $R(m)$ given by
\cite{hamana03}
\begin{equation}
\sigma_v(m,a)= \left[1+\delta(R_{\rm local})\right]^{2\mu(R_{\rm local})}aH(a)D_a \frac{d \ln D_a}{d \ln a}
\left(1-\frac{\sigma_0^4(m)}{\sigma^2_1(m)\sigma^2_{-1}(m)}\right)^{1/2}\sigma_{-1}(m,a)
\end{equation}
and \cite{sheth02}
\begin{equation}
\mu(R_{\rm{local}})\equiv 0.6 \sigma^2_0(R_{\rm{local}})/\sigma^2_0(10\,{\rm Mpc}/{\rm h}).
\end{equation}
Following \cite{hamana03}, $R_{\rm local}$ is obtained empirically using N-body simulations via the 
condition 
$\sigma_0(R_{\rm{local}})=0.5(1+z)^{-0.5}$.

Then the probability density function of the line-of-sight peculiar velocity component at some redshift $z$
is given by \cite{sheth02}
\begin{equation}
f(v,a)=\frac{\int  dm\,mn(m|\delta)p(v|m,\delta,a)}{\int dm\,mn(m|\delta,a)}
\label{fpdf}
\end{equation}
where $n(m|\delta) dm$ is the number density of halos that have mass between $m$ and $m+dm$ in
a region with overdensity $\delta$. The dependence of these
quantities on redshift is left implicit. 

Finally, in order to connect to a readily observable quantity, we write the fraction of clusters that have 
velocity between $v$ and $v+\delta v$ as 
\begin{equation}
n_v(v,\delta v,a)=\int_{\delta v}dv f(v,a).
\label{velpdf}
\end{equation}

\subsection{Mean Pairwise Velocity}

The mean relative peculiar velocity $v_{ij}(r)$ between all pairs of halos at comoving separation $r$  and  scale factor $a$
can be related to the linear two-point correlation function for dark matter using large-scale bias and the 
pair conservation equation \cite{davis77}:
\begin{equation}
v_{ij}(r,a)=-\frac{2}{3}H(a)a\frac{d \ln D_a}{d \ln a}\frac{r\bar{\xi}^{\rm halo}(r,a)}{1+\xi^{\rm halo}(r,a)}.
\label{v12}
\end{equation}
The two-point correlation function can be computed via
\begin{equation}
\xi^{\rm halo}(r,a)= \frac{D_a^2}{2\pi^2r} \int_0^{\infty} dk k \sin kr P(k)b_{\rm halo}^{(2)},
\end{equation}
while the two-point correlation function averaged over a sphere of radius $r$ can be written as
\begin{equation}
\bar{\xi}^{\rm halo}(r,a)= \frac{D_a^2}{2\pi^2r^2}\int_0^r dr r \int_0^{\infty} dk k \sin kr P(k)b_{\rm halo}^{(1)}
\end{equation}
where average halo bias factors are given by
\begin{equation}
b_{\rm halo}^{(q)} \equiv \frac{\int dm\,mn(m)b(m)^q W^2[kR(m)]}{\int dm\,mn(m)W^2[kR(m)]}.
\end{equation}

Direct evaluation of the above expression for mean pairwise peculiar velocity requires knowledge of 
all three velocity components for both halos. In practice, it is only possible to determine the radial velocity components,  so we need an estimator $v_{ij}^{\rm est}$ which depends only on the radial velocities.
Consider two clusters at positions ${\bf r}_i$ and ${\bf r}_j$ moving with velocities 
${\bf v}_i$ and ${\bf v}_j$. The radial components of velocities can be written as 
$v^{r}_i= {\hat{\bf r}}_i\cdot{\bf v}_i$ and  $v^{r}_j= {\hat{\bf r}}_j\cdot{\bf v}_j$. 
Following \citep{ferreira99}, 
$\langle v^{r}_i-v^{r}_j \rangle= v_{ij}^{\rm est}{\hat {\bf r}}\cdot[{\hat{\bf r}}_i+{\hat{\bf r}}_j]/2$ 
where ${\bf r}$ is the unit vector along the line joining the two clusters and
${\bf\hat r}$ is the unit vector in the direction ${\bf r}$. Then minimizing $\chi^2$ gives 
\begin{equation}
v_{ij}^{\rm est}= 2\frac{\Sigma(v^{r}_i- v^{r}_j)p_{ij}}{\Sigma p_{ij}^2}
\label{v12est}
\end{equation}
where $p_{ij} \equiv {\bf r}\cdot({\bf r}_i + {\bf r}_j)$ and the sums are over all pairs of
clusters with separation $r$.

\subsection{Velocity Correlation Function} 

In addition to the mean relative peculiar velocity between two halos, we can also
consider correlations of these velocities. Assuming statistical isotropy,
the only non-trivial correlations will be of the 
velocity components along the line connecting the clusters and of the velocity components
perpendicular to the line connecting the clusters; furthermore, these correlations will
only depend on the separation $r=|{\bf r}_i-{\bf r}_j|$.
Geometrically, the correlation of radial velocities must be of the form \citep{peel06}
\begin{equation}
\Psi_{ij}=\Psi_{\perp}\cos\theta + (\Psi_{\parallel}-\Psi_{\perp})
\frac{(r_i^2+r_j^2)\cos\theta-r_ir_j(1+\cos^2\theta)}{r_i^2+r_j^2-2r_ir_j\cos\theta}
\end{equation}
where $\theta = {\bf{\hat r}}_i\cdot{\bf{\hat r}}_j$ is the angle between the two cluster
positions; $\Psi_{\perp}(r)$ and $\Psi_{\parallel}(r)$ denotes the correlations perpendicular to the line of 
separation ${\bf r}$ and parallel to it, respectively.
Including the fact that high-density regions have lower rms velocities than random patches and allowing 
the two halos to have different masses, the expressions for correlations can be written as 
\citep{sheth01,gorski88}
\begin{equation}
\Psi_{\perp,\parallel}(m_i,m_j|r)= \frac{\sigma_0(m_i)\sigma_0(m_j)}{\sigma_{-1}(m_i)\sigma_{-1}(m_j)} 
a^2\frac{H(a)^2}{2\pi^2}\left[\frac{d \ln D_a}{d \ln a}\right]^2 D_a^2 
\int dk P(k) W[kR(m_i)]W[kR(m_j)]K_{\perp,\parallel}(kr) 
\end{equation}
where
\begin{equation}
K_{\perp}= \frac{j_1(kr)}{kr},\qquad\qquad K_{\parallel}= j_0(kr)-2\frac{j_1(kr)}{kr}
\end{equation}
with $j_0(kr)$ and $j_1(kr)$ the spherical Bessel functions.

With all the above ingredients, the correlation function for the velocity components
perpendicular to the line connecting the clusters can be written as
\begin{equation}
\langle v_iv_j\rangle_\perp(r,a)= \left[H(a)a
\frac{d \ln D_a}{d \ln a}D_a\right]^2\int dm_i \frac{m_i n(m_i)}{\bar \rho}
\int dm_j \frac{m_j n(m_j)}{\bar \rho}\frac{1+b(m_i)b(m_j)\xi(r)}{[1+\xi(r)]}\Psi_\perp(m_i,m_j|r)
\end{equation}
where ${\bar \rho}= \int dm m n(m)$.
Note that the above expression is a slight modification from Eq.~(23) of Ref.~\citep{sheth01}. The 
expression
for the correlation of the parallel velocity components is obtained simply by replacing $\Psi_\perp$
with $\Psi_\parallel$. Performing the ensemble average yields
\begin{equation}
\langle v_iv_j\rangle_\perp(r,a)= a^2H(a)^2\left(\frac{d \ln D_a}{d \ln a}\right)^2 D_a^2
\frac{1}{1+\xi^{\rm halo}(r,a)^2}
\frac{1}{\bar \rho^2}\left[I_1+\xi^{\rm halo}(r,a)I_2\right]
\label{v1v2perp}
\end{equation}
where 
\begin{equation}
I_1=\int  dk K_\perp(kr)P(k)\left[\int dm m n(m) \frac{\sigma_0(m)}{\sigma_{-1}(m)} W[kR(m)]\right]^2,
\label{I1}
\end{equation}
\begin{equation}
I_2=\int dk K_\perp(kr)P(k)\left[\int dm m n(m)b(m)\frac{\sigma_0(m)}{\sigma_{-1}(m)}W[kR(m)]\right]^2.
\label{I2}
\end{equation}

Although the above expression holds for both the parallel and perpendicular components, in simulations  
$\Psi_{\parallel}$ is mostly negative or zero due to the heavy influence of infall at large separations 
\citep{peel06}. However, this anticorrelation is not seen in linear perturbation theory or in the halo
model, which both predict positive correlation for pair comoving separations less than  
40 Mpc; for separations larger than 40 Mpc, the theory and simulations are consistent, but
the parallel component correlation is essentially zero. Given this discrepancy between known analytical
models and simulations for the parallel correlation function in the region where the
signal is non-negligible, we only consider 
$\langle v_iv_j\rangle_\perp(r,z)$ in the rest of this paper.

\section{Comparison with Simulations}
\label{sec:simulations}

The statistics computed in the previous section are based on the halo model of structure formation
combined with linear perturbation theory. Since galaxy clusters are rare objects and their distribution 
can be described well in the quasi-linear regime of structure formation, we expect that these 
approximations for velocity statistics should be reasonably accurate. 
Here we verify that they are good approximations
to the actual galaxy cluster velocity statistics extracted from the
the VIRGO dark matter simulation \citep{virgo}.
We use the octant sky survey (PO) lightcone output of LCDM cosmology, with $\sigma_8=0.9$, $n_s=1$, 
$\Omega_m=0.3$, $\Omega_{\Lambda}=0.7$ and $h=0.7$. The maximum redshift of the light cone is 
$z_{\rm max}= 1.46$ and the radius of extent is $R_{\rm max}=3000$ Mpc/h. 
The data is binned in redshift slices of width $\delta z=0.2$
from $z=0$ to $z=1.4$ 

The statistics defined in the previous section apply to infinitesimal intervals
in redshift. When comparing with data binned in redshift, it is necessary to normalize the
velocity statistics properly to reflect this binning. We do this by averaging the
above theoretical expressions for the statistics over a given bin in $z$ to obtain
a binned estimator of the underlying statistic. Additionally, for the case of the velocity
probability distribution function, a realistic measurement will provide numbers
of clusters in a set of line-of-sight velocity bins. In this case, the relevant statistic
for comparison becomes the theoretical probability that the velocity of a given
cluster is in a particular velocity bin; the above expression for probability density in
infinitesimal velocity bins must be integrated over the width of the velocity bin. This
gives the correct relative probabilities between any two velocity bins, but all must
then be renormalized by a constant factor to enforce the condition that the sum
of the probabilities for all bins be unity. All comparisons with simulations
below use these binned versions of the underlying statistics defined in
the previous section. 

\begin{figure}
\resizebox{85mm}{!}{\includegraphics{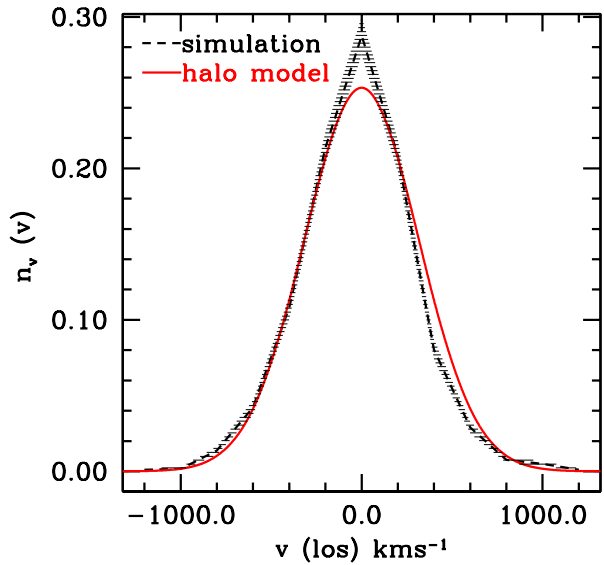}}
\caption{A comparison between the probability distribution function $n_v$
evaluated directly using the Virgo lightcone numerical simulation (dotted curve with error bars)
and approximated using the analytic halo model formula, Eq.~(\ref{velpdf}) (solid red curve).
Error bars are Poisson plus cosmic variance errors for one octant sky coverage.}
\label{pdfcomparefig}
\end{figure}

\begin{figure}
\resizebox{85mm}{!}{\includegraphics{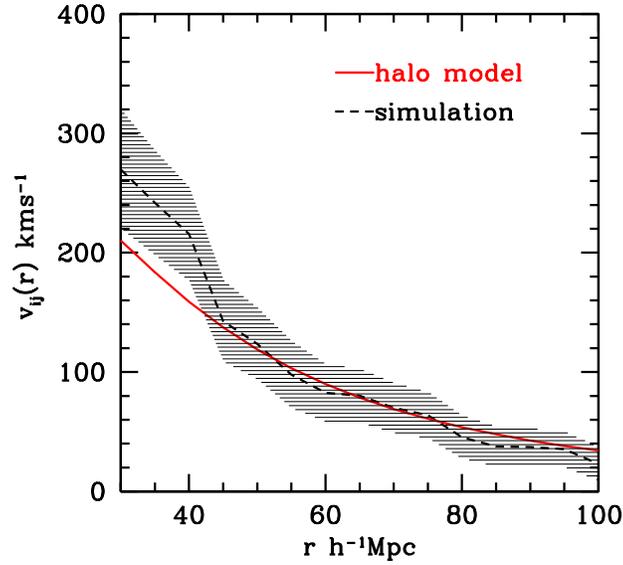}}
\caption{A comparison between the mean pairwise streaming velocity $v_{ij}(r)$
evaluated directly using the Virgo lightcone numerical simulation (dashed line with 1$\sigma$ errors 
given by the blue dotted lines)
and approximated using the analytic halo model formula, Eq.~(\ref{v12}) (red solid curve). 
The error range includes Poisson
and cosmic variance errors for one octant sky coverage, plus random measurement errors 
of 100 km/s.}
\label{v12comparefig}
\end{figure}

\begin{figure}
\resizebox{85mm}{!}{\includegraphics{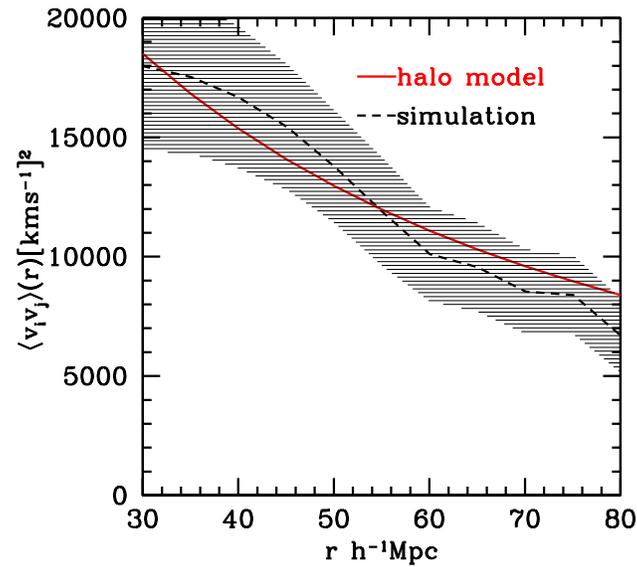}}
\caption{Same as Figure~\ref{v12comparefig} except for the velocity correlation
$\langle v_iv_j\rangle_\perp(r)$ and the analytic formula Eq.~(\ref{v1v2perp}).}
\label{v1v2perpcomparefig}
\end{figure}

Figure~\ref{pdfcomparefig} shows $n_v$ in the redshift slice between $z=0$ and $z=0.2$ both from the 
simulation and using Eq.~(\ref{velpdf}) for the velocity probability distribution function. The analytical 
model agrees fairly well with the simulation; the error bars denote the $1\sigma$ errors including Poisson 
error and errors due to cosmic variance. Error modeling is discussed in detail in the next Section. Note 
that the error bars shown in Figure~\ref{pdfcomparefig} are for a large
future 5000 square degree velocity survey (one octant of the sky). 
Figures~\ref{v12comparefig} and \ref{v1v2perpcomparefig} compare the simulation  with
Eq.~(\ref{v12}) for the mean pairwise streaming velocity (using the estimator Eq.~(\ref{v12est})) 
and Eq.~(\ref{v1v2perp}) for the velocity correlation function, respectively. The plots shows that the halo 
model agrees well with the simulated data at separations greater than 30 Mpc/h for velocity
correlation, and greater than 40 Mpc/h for mean pairwise streaming velocity with a discrepancy somewhat
larger than $1\sigma$ for $r$ between 30 and 40 Mpc/h.  For the velocity probability distribution function, 
we find a good fit when the velocity data is smoothed over a scale of 10 Mpc. The smoothing on this scale 
reduces the effect of nonlinear physics which is difficult to model semianalytically.

\begin{figure}
\resizebox{85mm}{!}{\includegraphics{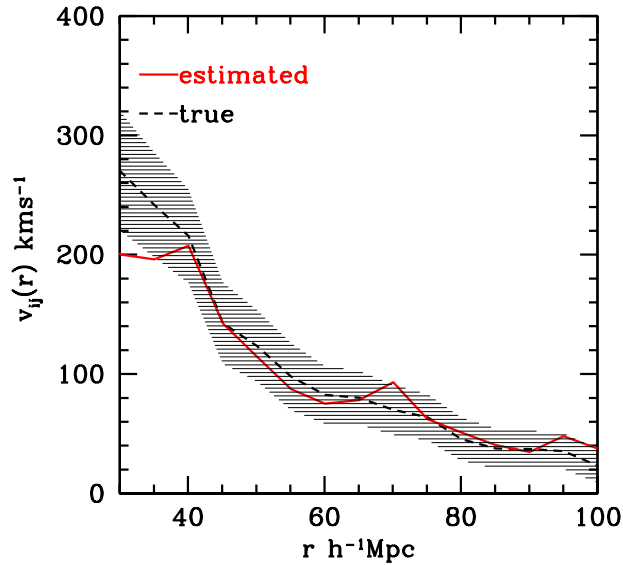}} 
\caption{The solid red line shows $v_{ij}^{\rm est}$ computed from the Virgo simulation using only the 
radial velocities, Eq.~(\ref{v12est}), while the dashed line shows $v_{ij}$ and shaded 1$\sigma$ errors computed using all three velocity components, the same as in Fig.~\ref{v12comparefig}}. 
\label{radialfullcomparefig}
\end{figure}

Figure~\ref{radialfullcomparefig} displays a comparison between the estimated mean pairwise streaming 
velocity $v_{ij}^{\rm est}$ obtained only from the radial component of velocity using Eq.~(\ref{v12est}) and 
the full $v_{ij}$ obtained from all three components of velocity in the simulation. For an ideal estimator, 
these quantities would be
exactly the same; the actual estimator in general does quite well, except for a $1\sigma$ discrepancy at 
separations below 30 Mpc/h. The error range is the same as for Figure~\ref{v12comparefig}. 

\section {Error Sources}
\label{sec:errors}

Measurement of the radial velocity of individual clusters via their kinematic Sunyaev-Zeldovich
signal is affected by various error sources, including detector noise in the microwave maps,
separating the small signal
from other larger signals at the same frequencies (particularly the thermal SZ signal,
infrared point sources, and gravitational lensing by the cluster), the internal velocity
dispersion of the intracluster medium, and X-ray temperature measurement errors. In this Section,
we call the
total error from all of these sources ``measurement error." We also consider separately
the errors arising from cosmic variance and Poisson noise; 
both of these error sources are independent of the measurement
errors for any individual cluster. 

\subsection{Velocity Measurement Errors}

Upcoming multi-frequency Sunyaev-Zeldovich measurements with arcminute resolution and
few $\mu$K sensitivity have the potential to obtain galaxy cluster peculiar velocities. However, the 
kinematic Sunyaev-Zeldovich signal is small compared to the thermal SZ signal, and is
spectrally indistinguishable from the primary microwave blackbody fluctuations or their gravitational 
lensing. In addition, radio and infrared galaxies contribute substantial signal in the microwave bands, and 
are expected to be spatially correlated with galaxy cluster positions \cite{coble07}. 
Comparatively modest error sources can substantially hinder cluster velocity measurements if
they are not well understood and accounted for. 

Major potential sources of  error in measuring the velocities of individual galaxy clusters include
internal cluster gas velocities, the confusion-limited noise from point sources, uncertainties
in extrapolating measured point sources to the frequencies of a particular experiment,
instrumental noise, and the particular frequency bands available. 
Previous studies shows that primary microwave background fluctuations plus point sources set a 
confusion limited velocity error of around 200 km/s for an experiment with arcminute resolution and few 
$\mu$K sensitivity \citep{knox04, aghanim01, tegmark96}, provided no other point source follow-up 
observations are utilized. The bulk flow of the gas in the intracluster medium contributes to an 
irreducible error of 100 to 150 km/s \citep{nagai03, diaferio05}. Also, Ref.~\citep{sehgal05} shows that to 
extract velocity from SZ observations at the three ACT frequency channels (145, 220, and 280 GHz), a 
followup measurement of X-ray temperature of the cluster is needed to break a spectrum degeneracy 
between cluster gas velocity, optical depth, and temperature. While Ref.~\cite{diaferio05} studied over 100 
simulated clusters, the rest of these studies use only a few. All of these error sources require
detailed simulations of particular experiments observing realistic simulated clusters and optimal
algorithms for extracting cluster velocities from measurements in particular frequency bands
and at given instrumental noise levels. The ultimate distribution of velocity errors is still
uncertain and future study in this direction is needed. In order to study the effect of
measurement errors on parameter estimation, we make the simple assumption that
velocity errors have a normal distribution with a magnitude between 100 and 500 km/s. 
Directly adding all of the known sources of error from previous studies gives velocity
measurement errors typically in the range of 400 to 500 km/s; however, with further
understanding of systematic errors and point sources, the error budget may be 
reduced. 

\subsection{Redshift Errors}

In addition to cluster velocity, we must measure cluster redshift to construct
the estimators of the mean pairwise velocity and the velocity correlation, which
involve knowledge of the separation vector between the two clusters. For clusters
at cosmological distances, the Hubble contribution to its redshift will typically be
much larger than its peculiar velocity contribution, which we can also correct for
with a direct velocity measurement, so direct error in the cluster redshift will
be the largest contributor to the cluster position error. Typically, we will
be concerned with cluster separations larger than 30 Mpc/h, for which
the cluster velocity field is in the mildly nonlinear regime and can be
well described by the halo model approximation. 


A redshift error of 500 km/sec corresponds to
a direct Hubble distance error of around 5 Mpc/h, typically only 25\% of
the closest cluster separation of interest; even for redshift errors of 1000 km/sec,
most pair separations will not be dominated by this error. For the remainder of
this paper, we assume that the cluster sample for which velocities are determined
also have spectroscopic redshifts from which their distances are determined, and
we assume that the distance error effect on the cosmological parameters will
be negligible compared to the direct velocity errors. For spectroscopic measurements
of many galaxy clusters, the distance to lowest order is simply determined by the average
of the galaxy redshifts, with an error given roughly by the cluster galaxy velocity
dispersion divided by the square root of the number of clusters' galaxies. Cluster line-of-sight
velocity dispersions will typically be 500 km/sec, so multi-object
spectroscopy can clearly provide adequate redshift measurements. The systematic
error induced because not all clusters will be virialized is potentially important,
although beyond the scope of this paper. 

Spectroscopic redshifts for a galaxy cluster at $z=1$ requires roughly an hour
of observation on an 8-m class telescope. Spectroscopic follow-up of
hundreds of clusters per year is a large program for a single telescope;
spectroscopic redshifts for thousands of clusters will comprise a multi-year
program on more than one telescope. This is likely to be a significant portion of
the effort and expense in building a cluster peculiar velocity survey with thousands
of clusters. Note that cluster galaxy spectroscopic redshifts are also valuable
for dynamical mass estimates; see, e.g., \cite{diaferio_dyn,diaferio_dyn1}. 
The ACT collaboration has plans for spectroscopic follow-up
observations of SZ-detected clusters using the Southern African Large Telescope (SALT),
a new 10-meter class instrument. If only photometric redshifts are available,
typically giving a distance accuracy of one to two percent times $1+z$, cosmological
constraints must be re-evaluated. In general, constraints will be less stringent, although it is
not immediately clear whether the resulting distance errors will have an effect
which is significant compared to  the velocity errors. In our case, redshift errors propagate only into the 
geometric portions of the mean pairwise streaming velocity
and velocity correlation estimators, but the velocity errors are unaffected. 
This issue will be addressed in detail elsewhere.

\subsection{Cosmic Variance and Poisson Noise}

In addition to measurement errors for individual cluster velocities, cosmological quantities are
also subject to errors from cosmic variance (any particular region observed may have different
statistical properties from the average of the entire universe) and Poisson 
errors due to the finite size of the cluster velocity sample used to estimate the velocity statistics. 
Here we discuss these errors
for each of the three velocity statistics. Detailed derivations of the expressions in the rest
of this Section are given in the Appendices.

\subsubsection{Probability Density Function}

Consider a cluster velocity survey with a measured redshift for each cluster. 
For the probability density function, we write cosmic covariance between two different velocity--redshift bins $[v,z]_i$ and 
$[v,z]_j$ as $C^{n_v}_{ij}$, which can be expressed as 
\begin{equation}
C^{n_v}(ij)= \frac{3D_{a_i}D_{a_j}}{R_\Omega}n_in_j \int dk k^2P(k)j_1(kR_{\Omega})
\label{C_fij}
\end{equation}
where 
\begin{equation}
n_v(v,z)= \frac{\int dm mb(m,a){\bar n(m)}p(v|m,\delta,a)}{\int dm m {\bar n(m)}}
\end{equation}
and $R_{\Omega}$ is the comoving length of the redshift bin within the sky
survey region \cite{hogg00}.

For Poisson errors, let $N_i$ be the total number of clusters in bin $i$. We are interested in
the error in $n_i = N_i/N_z$ with $N_z$ the total number of clusters in a particular
redshift bin summed over all velocities; the measured $n_i$ corresponds to the theoretical
quantity $n_v(v,z)$, Eq.~(\ref{velpdf}), integrated over the velocity--redshift bin $[v,z]_i$. The 
expression for Poisson errors can be written as 
\begin{equation}
\delta n_i= ({\sqrt n_i} + n_i)/{\sqrt N_z}
\end{equation}
where the first term is from the error in $N_i$ and the second from the error in $N_z$. 

Random velocity measurement errors will smear out the velocity PDF.  We 
quantify the effect of measurement errors by convolving the PDF with a normal distribution of velocity 
errors,
\begin{equation}
n_v^{\rm obs}(v,\delta v, z)= \int_{\delta v} dv \int_{v_l}^v dv' f(v',z)\exp[-(v'-v)^2/{2\sigma_v^2}]
\label{nverror}
\end{equation}
where  $\sigma_v$ is the dispersion of the normally distributed velocity errors and the integral
is over the velocity bin.
Then the expression for the total covariance  can be written as 
\begin{equation}
C^{n_v}_t(v_i, z_i;v_j, z_j)=C^{n_v}(ij)+(\delta n_i)^2\delta_{ij}
\label{C_f_total}
\end{equation}

\begin{figure}
  \begin{center}
    \begin{tabular}{cc}
        \resizebox{85mm}{!}{\includegraphics{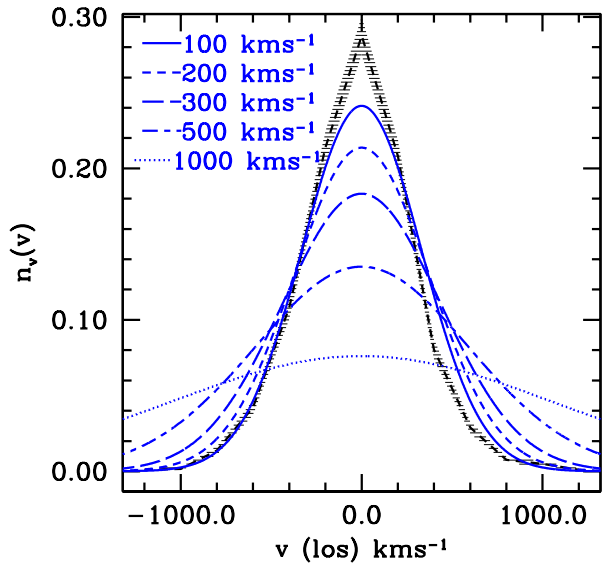}} \\
    \end{tabular}
    \caption{The effect of measurement errors on the velocity probability distribution function: from top
to bottom, velocity measurement errors of $\sigma_v=$100, 200, 300, 500, and 1000 km/s. Also shown 
are the probability distribution function evaluated directly using the Virgo lightcone numerical simulation 
(dotted curve with error bars) from Figure~\ref{pdfcomparefig}}
    \label{pdf_vs_v}
  \end{center}
\end{figure}

The various curves in Fig.~\ref{pdf_vs_v}  show the effect of random velocity errors of
different sizes, Eq.~(\ref{nverror}), while the top dotted curve with shaded error region gives the
actual value for the probability distribution function from the VIRGO simulation with Poisson
plus cosmic variance errors. Smearing the distribution by random velocity errors is 
largely degenerate with the effect of varying cosmological parameters. This means that the velocity 
probability distribution function  as a probe of cosmology  is limited by how well the measurement error 
can be understood from simulated measurements.

\subsubsection{Mean Pairwise Streaming Velocity}

The mean pairwise streaming velocity statistic is binned in pair separation and redshift.
The cosmic covariance between two bins $[r,z]_p$ and $[r,z]_q$ can be written as 
\begin{equation}
C^{v_{ij}}(pq)=\frac{32\pi}{9V_\Omega}\frac{H(a_p)a_p}{1+\xi^{\rm halo}(r_p,a_p)}
\frac{H(a_q)a_q}{1+\xi^{\rm halo}(r_q,a_q)}
\left(\frac{d \ln D_a}{d \ln a}\right)_{a_p}\left(\frac{d \ln D_a}{d \ln a}\right)_{a_q}
\int dk k^2 |P(k)|^2j_1(kr_p)j_1(kr_q).
\label{C_vij_cosmic}
\end{equation}
We add in quadrature the Poisson error and measurement error for $n_{\rm pair}$ cluster pairs
and write the total covariance as
\begin{equation}
C^{v_{ij}}(r_p,z_p;r_q,z_q)=C^{v_{ij}}_{\rm cosmic}(pq)
+\left(\frac{v_{ij}^2}{n_{\rm pair}}+\frac{2\sigma_v^2}{n_{\rm pair}}\right)\delta_{pq}
\label{C_vij}
\end{equation}

\begin{figure}
  \begin{center}
    \begin{tabular}{cc}
      \resizebox{85mm}{!}{\includegraphics{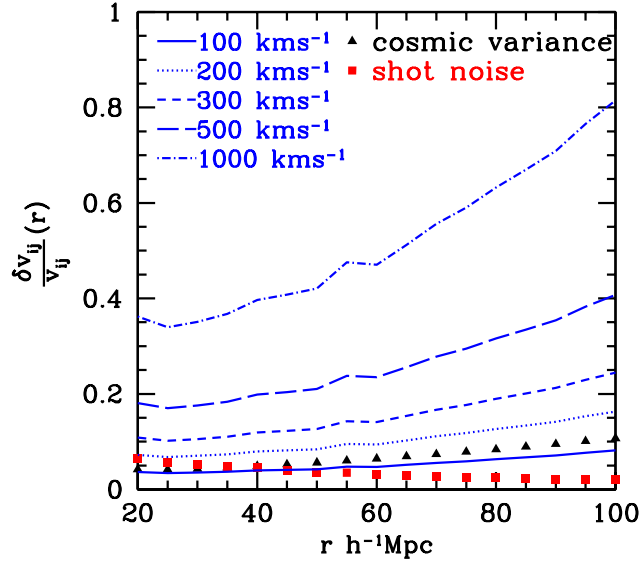}} 
    \end{tabular}
    \caption{Fractional errors $\delta v_{ij}/v_{ij}$ for a cluster velocity survey covering 5000 square
    degrees: the red square points represents the Poisson error; black triangles represents cosmic variance 
    and the Blue lines represents measurement errors (from bottom to top $\sigma_v$=100, 200, 300, 500 
    and 1000 km/s). Note that all the errors scales as {$\sqrt f_{sky}$} for other survey areas.}
    \label{fracerrors_vij}
  \end{center}
\end{figure}

Figure~\ref{fracerrors_vij} plots fractional errors for $v_{ij}$ as a function of pair separation for a survey 
area of 5000 deg$^2$. For a survey area $f_{\rm sky}$, fractional errors scales as roughly 
$\sqrt{f_{\rm sky}}$. Note that the Poisson error decreases for larger separation since more clusters pairs 
are available to average over, whereas cosmic variance has an increasing effect at larger separation. The 
combined effect of cosmic variance plus Poisson errors dominates the error budget when velocity 
measurement errors are below 200 km/s. Note that even when the measurement errors are as high as 
$\sigma_v=500$ km/s, the total error is typically 50\% of the magnitude of mean pairwise streaming 
velocity. We will show in Sec.~\ref{sec:constraints} that this fact makes mean pairwise streaming velocity a 
potentially useful probe to study cosmology.  

\subsubsection{Velocity Correlation Function}

Similarly for the velocity correlation function, the expression for cosmic covariance can be written as 
\begin{eqnarray}
C^{\langle v_iv_j\rangle}_{\rm cosmic}(pq)&=& \frac{8\pi}{V_{\Omega}{\bar \rho}^2(p) {\bar \rho}^2(q)}
\left[\frac{d \ln D_a}{d \ln a}\right]^2_{a_p}\left[\frac{d \ln D_a}{d \ln a}\right]^2_{a_q}
\frac{a_p^2D^2_{a_p}H^2(a_p)}{1+\xi^{\rm halo}(r_p,a_p)}
\frac{a_q^2D^2_{a_q}H^2(a_q)}{1+\xi^{\rm halo}(r_q,a_q)}\nonumber \\
&&\qquad\qquad\qquad\qquad\times 
\int dk j_1(kr_p)j_1(kr_q)[P(k)]^2\langle p\rangle^2_m\langle q\rangle^2_m
\label{C_vivj_cosmic}
\end{eqnarray}
using the notational abbreviation 
\begin{equation}
\langle x\rangle_m\equiv\int dm\,m \frac{dn}{dm} W(kR(m))\frac{\sigma_0(m)}{\sigma_{-1}(m)} \,x.
\label{m-avg}
\end{equation}
In Eq.~(\ref{C_vivj_cosmic}) we have ignored the contribution of the second ($I_2$) term in 
Eq.~(\ref{v1v2perp}). At larger separations relevant here, this term, being weighted by $\xi(r)$, 
is an order of magnitude smaller than the first term and hence has negligible contribution to the 
cosmic variance.

Again we add in quadrature the Poisson error and measurement error for $n_{\rm pair}$ cluster pairs
and write the total covariance as
\begin{equation}
C^{\langle v_iv_j\rangle}_t[r_p,z_p|r_q,z_q]= C^{\langle v_iv_j\rangle}(pq)
+\left[\frac{\langle v_iv_j\rangle(r,z)}{\sqrt{ n^{pair}(r,z)}}\right]^2
+\left[\frac{1}{n^{pair}} \Sigma [\delta (v^2)+ (\delta v)^2]\right]^2
\label{C_vivj}
\end{equation}

\begin{figure}
  \begin{center}
    \begin{tabular}{cc}
       \resizebox{85mm}{!}{\includegraphics{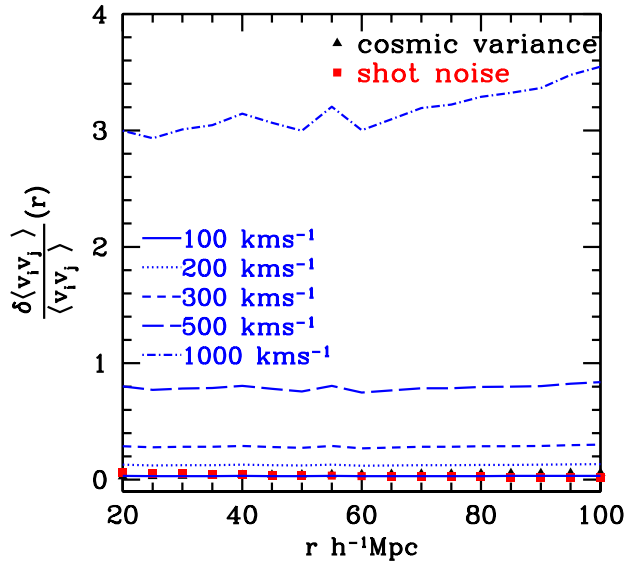}}
    \end{tabular}
    \caption{Same as in Figure~\ref{fracerrors_vij} for the fractional error 
    $\delta(\langle v_iv_j\rangle)/\langle v_iv_j\rangle$.}
    \label{fracerrors_vivj}
  \end{center}
\end{figure}

Figure~\ref{fracerrors_vivj} shows the various errors in the velocity correlation function. The trends are 
similar to those for mean pairwise streaming velocity. Measurement errors dominate the error budget for 
$\sigma_v>200$ km/s. Note however the increase in fractional errors with the increase in measurement
errors. For $\sigma_v= 500$ km/s, the contribution of measurement errors to the total error is 
almost 90\%, nearly double that for the case of mean pairwise streaming velocity. 

\section{Constraints on Dark Energy Parameters}
\label{sec:constraints}

Now we consider constraints on dark energy parameters for various survey areas and over a range of 
velocity errors. Following the Dark Energy Task Force, we describe the dark energy 
in terms of three phenomenological parameters: its current energy density $\Omega_\Lambda$,
and two parameters $w_0$ and $w_a$ describing the redshift evolution of its equation
of state $w(a) = w_0 + (1-a)w_a$. 
Assuming a spatially flat universe, the set of cosmological parameters ${\bf p}$ on which the velocity field 
depends are the normalization of the matter power spectrum $\sigma_8$ (or equivalently
the normalization constant $B$ in Eq.~(\ref{transferfunction})), the power law index of the primordial power 
spectrum $n_S$, and the Hubble parameter $h$, plus  
the dark energy parameters. 
We perform a simple Fisher matrix analysis to find constraints on these parameters
from measurements of the three velocity statistics described in Sec.~\ref{sec:statistics}.   

We consider a  fiducial model similar to that assumed in the DETF report \citep{detf} with 
$\sigma_8=0.9$, $n_S=1$, $h=0.7$, $\Omega_\Lambda= 0.72$, $w_0= -1$, $w_a= 0$. 
To make quantitative comparisons with the conclusions of the DETF report, we compute
values for the expression $[\sigma(w_0)\sigma(w_p)]^{-1}$, which is listed in the DETF
summary tables. We refer to this as the ``Figure of Merit'' (FOM) for convenience, although
this term refers to a slightly different quantity (inverse area of the ellipse of $95\%$ confidence limit in the 
$w_p-w_a$ plane) in the DETF report. Here $w_p$ is the equation of state at the pivot point defined as 
$w_p=w_0+(1-a_p)w_a$ with $a_p= 1+[F^{-1}]_{w_0w_a}/[F^{-1}]_{w_aw_a}$ and $F$ the Fisher 
information matrix for a given experiment. 

\begin{figure*}
  \begin{center}
       \resizebox{7 in}{!}{\includegraphics{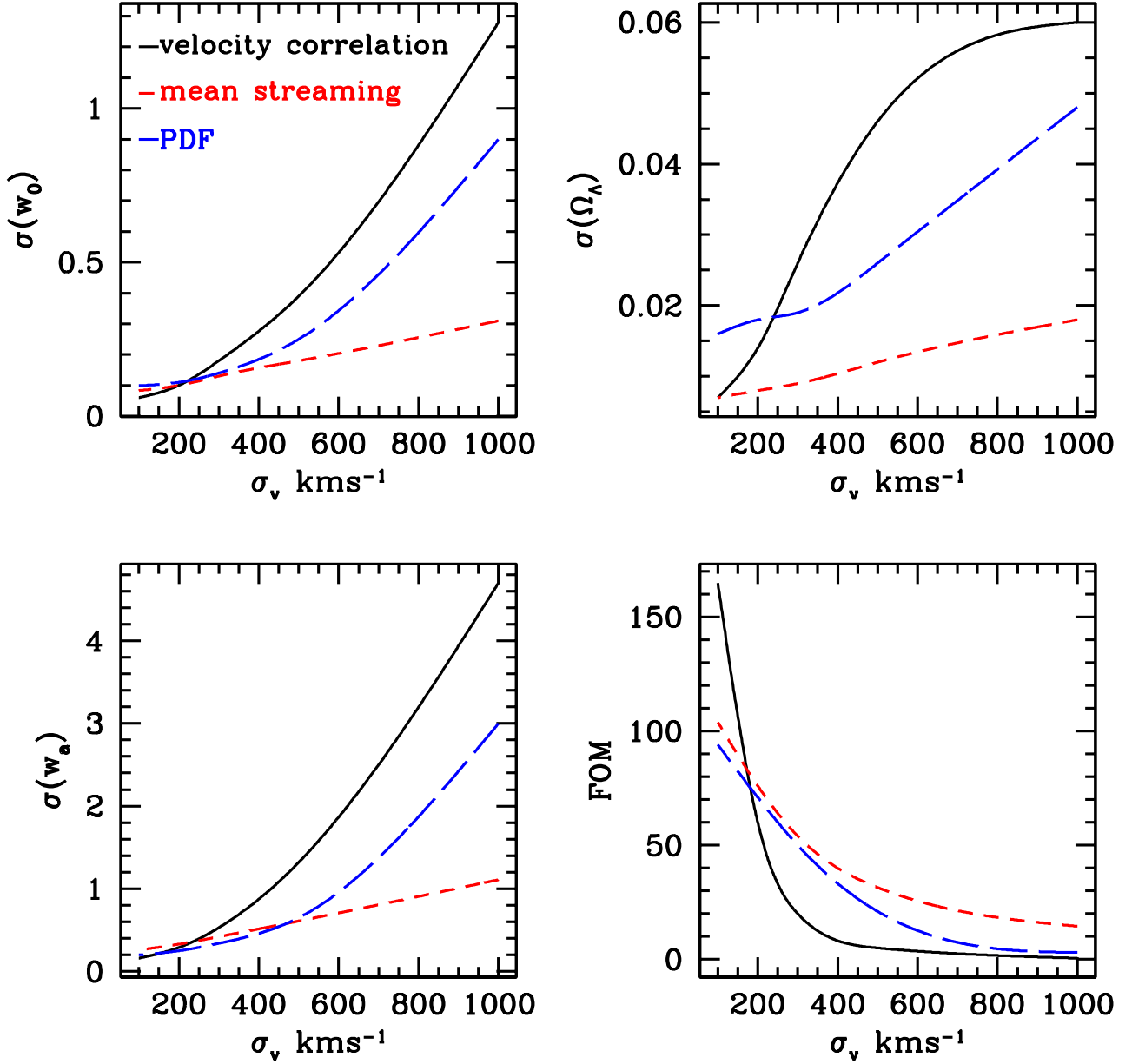}}
    \caption{The change in  1$\sigma$ parameter constraints  with velocity error (normal distribution of 
    width $\sigma_v$) for a 4000 deg$^2$ survey area, for the three statistics $n_v$ (blue dashed), $v_{ij}$ 
    (red short dashed) and $\left\langle v_iv_j\right\rangle$ (black solid). The
    four panels are for the parameters  $w_0$ (top left),
    $\Omega_\Lambda$ (top right),  $w_a$ (bottom left), and the Figure of Merit (bottom right).}
    \label{p_vs_v}
  \end{center}
\end{figure*}

\begin{table*}
\caption{1$\sigma$ errors on dark energy parameters for a 4000 deg$^2$ survey area plus cosmological 
priors from  Planck and HST \citep{detf, hst}, assuming a spatially flat cosmology.}
\begin{center}
\begin{tabular}{ccccccccccccc}
\hline
$\sigma_v$ &{ }&$w_0$&{ }&{ }&$w_a$&{  }&{ }&$\Omega_{\Lambda}$&{ } &{ }&$[\sigma(w_0)\sigma(w_p)]^{-1}$&{ }\\
&${\langle v_iv_j\rangle}$&${v_{ij}}$&${n_v}$&${\langle v_iv_j\rangle}$&${v_{ij}}$&${n_v}$&${\langle v_iv_j\rangle}$&${v_{ij}}$&${n_v}$&${\langle v_iv_j\rangle}$&${v_{ij}}$&${n_v}$\\\hline
100 & 0.06 & 0.083 & 0.099 & 0.16 & 0.26 & 0.2 & 0.007 & 0.007 & 0.016 & 165 & 104 & 94 \\
200 & 0.1 & 0.1 & 0.11 & 0.29 & 0.33 & 0.25 & 0.014 & 0.008 & 0.018 & 60 & 76 & 71 \\	
300 & 0.18 & 0.13 & 0.14 & 0.53 & 0.42 & 0.34 & 0.026 & 0.009 & 0.019 & 20 & 54 & 50 \\
500 & 0.39 & 0.18 & 0.25 & 1.32 & 0.61 & 0.65 & 0.046 & 0.012 & 0.026 & 5 & 31.5 & 21\\
1000 & 1.28 & 0.31 & 0.9 & 4.7 & 1.11 & 3.0 & 0.060 & 0.018 & 0.048 & 0.5 & 14.5 & 3.0\\\hline
\end{tabular}
\end{center}
\label{tab:fbg4000}
\end{table*}

\begin{table*}
\caption{Same as Table \ref{tab:fbg2000}, for a 2000 deg$^2$ survey area.}
\begin{center}
\begin{tabular}{ccccccccccccc}
\hline
$\sigma_v$ &{ }&$w_0$&{ }&{ }&$w_a$&{  }&{ }&$\Omega_{\Lambda}$&{ } &{ }&$[\sigma(w_0)\sigma(w_p)]^{-1}$&{ }\\
&${\langle v_iv_j\rangle}$&${v_{ij}}$&${n_v}$&${\langle v_iv_j\rangle}$&${v_{ij}}$&${n_v}$&${\langle v_iv_j\rangle}$&${v_{ij}}$&${n_v}$&${\langle v_iv_j\rangle}$&${v_{ij}}$&${n_v}$\\\hline
100 & 0.08 & 0.12 & 0.12 & 0.26 & 0.41 & 0.28 & 0.011 & 0.010 & 0.018 & 80 & 53 & 59 \\
200 & 0.13 & 0.14 & 0.15 & 0.43 & 0.51 & 0.35 & 0.011 & 0.011 & 0.020 & 31 & 39 & 47 \\	
300 & 0.25 & 0.18 & 0.19 & 0.77 & 0.63 & 0.47 & 0.035 & 0.013 & 0.022 & 11 & 29 & 32 \\
500 & 0.52 & 0.25 & 0.33 & 1.83 & 0.89 & 0.9 & 0.052 & 0.016 & 0.032 & 3 & 18 & 13 \\
1000 & 1.8 & 0.42 & 1.26 & 6.7 & 1.48 & 4.2 & 0.061 & 0.022 & 0.061 & 0.75 & 7.9 & 1.6 \\\hline
\end{tabular}
\end{center}
\label{tab:fbg2000}
\end{table*}

\begin{table*}
\caption{Same as Table \ref{tab:fbg4000}, for a 400 deg$^2$ survey area.}
\begin{center}
\begin{tabular}{ccccccccccccc}
\hline
$\sigma_v$ &{ }&$w_0$&{ }&{ }&$w_a$&{  }&{ }&$\Omega_{\Lambda}$&{ } &{ }&$[\sigma(w_0)\sigma(w_p)]^{-1}$&{ }\\
&${\langle v_iv_j\rangle}$&${v_{ij}}$&${n_v}$&${\langle v_iv_j\rangle}$&${v_{ij}}$&${n_v}$&${\langle v_iv_j\rangle}$&${v_{ij}}$&${n_v}$&${\langle v_iv_j\rangle}$&${v_{ij}}$&${n_v}$\\\hline
100 & 0.13 & 0.20 & 0.2 & 0.45 & 0.72 & 0.51 & 0.019 & 0.015 & 0.023 & 30 & 22 & 29 \\
200 & 0.24 & 0.25 & 0.24 & 0.76 & 0.92 & 0.64 & 0.034 & 0.017 & 0.026 & 11 & 16 & 21 \\	
300 & 0.41 & 0.31 & 0.31 & 1.39 & 1.15 & 0.85 & 0.048 & 0.020 & 0.031 & 4.0 & 11 & 14 \\
500 & 0.92 & 0.6 & 0.53 & 3.4 & 1.66 & 1.53 & 0.058 & 0.024 & 0.044 & 1.4 & 0.7 & 5.2\\
1000 & 3.6 & 0.78 & 2.42 & 13.3 & 3.0 & 8.0 & 0.061 & 0.033 & 0.061 & 0.38 & 3.3 & 0.7\\\hline 
\end{tabular}
\end{center}
\label{tab:fbg400}
\end{table*}

The Fisher information matrix for each of the three statistics is \cite{bk06}
\begin{equation}
F_{\alpha\beta}= \sum_{i,j}\frac{\partial \phi(i)}{\partial p_\alpha}[C^{\phi}_t(ij)]^{-1}\frac{\partial \phi(j)}{\partial p_\beta}
\end{equation}
where $\phi$ stands for either $n_v$, $v_{ij}(r,z)$ or $\langle v_iv_j\rangle(r,z)$,
$C^{\phi}(ij)$ is the total covariance matrix in each bin for the statistic $\phi$, 
Eqs.~(\ref{C_f_total}), (\ref{C_vij}), and (\ref{C_vivj}),  and the partial derivatives are evaluated
for the fiducial values of the cosmological parameters.
The values $i$ and $j$ index the bins $[r,z]_i$ and $[r,z]_j$ for the mean pairwise streaming velocity 
and velocity correlation function, while for $\phi= n_v$, $i$ and $j$ refer to $[v,z]_i$ and $[v,z]_j$. The inverse of the Fisher matrix
has diagonal elements which are estimates for the variances of each cosmological parameter 
marginalized over the values of the other parameters, and the non-diagonal elements give the
correlations between parameters. 

Figure~\ref{p_vs_v} shows the degradation of parameter constraints with increasing velocity error  
$\sigma_v$ for a 4000 deg$^2$ survey area. It is evident that parameter constraints from $v_{ij}$ are more
robust to increases in velocity error than those from  $n_v$ and $\langle v_iv_j\rangle$. This is because $
\delta v_{ij}$ depends linearly on $\sigma_v$, while $\delta \langle v_iv_j\rangle$ varies as 
$\sigma_v^2$ and for $n_v$ the distribution gets smeared with increases in $\sigma_v$. Constraints on 
$w_0$, $w_a$ and $\Omega_{\Lambda}$ change roughly by a factor of two and the constraint on the 
FOM by a factor of three,  for the factor of five increase in $\sigma_v$ from 200 to 500 km/s. Compare this
to the corresponding change for $\langle v_iv_j\rangle$: $w_0$, $w_a$ and $\Omega_{\Lambda}$ 
constraints change roughly by a factor of 6 to 8 and the FOM constraint by a factor of 30 for a similar 
change in $\sigma_v$. For $n_v$, the corresponding degradation in constraints are roughly by a factor 
1.5 to 3 for $w_0$, $w_a$ and $\Omega_{\Lambda}$ while the FOM constraint degrades by roughly a 
factor of 4. Table \ref{tab:fbg4000}  lists the constraints as a function of velocity error for a 4000 deg$^2$ 
survey area, while Tables \ref{tab:fbg2000} and \ref{tab:fbg400} give constraints for  2000 deg$^2$ and 
400 deg$^2$ respectively. 

Note that the velocity correlation function $\langle v_iv_j\rangle$ provides the best constraints on the dark 
energy equation of state ($w_0$, $w_a$, and FOM) for $\sigma_v<200$ km/s. It might be possible to 
achieve such values of errors in future surveys with better understanding of point source contamination 
and other systematics. However for more realistic near-term errors of  500 km/s, the mean pairwise 
peculiar velocity $v_{ij}$ provides better constraints on dark energy parameters, and this statistic
will be used in the following sections which consider how cosmological constraints will
be improved by using cluster velocity information.

\section{Complementarity of Cluster Velocities with Cluster Number Counts}

\begin{figure}
  \begin{center}
    \begin{tabular}{cc}
       \resizebox{85mm}{!}{\includegraphics{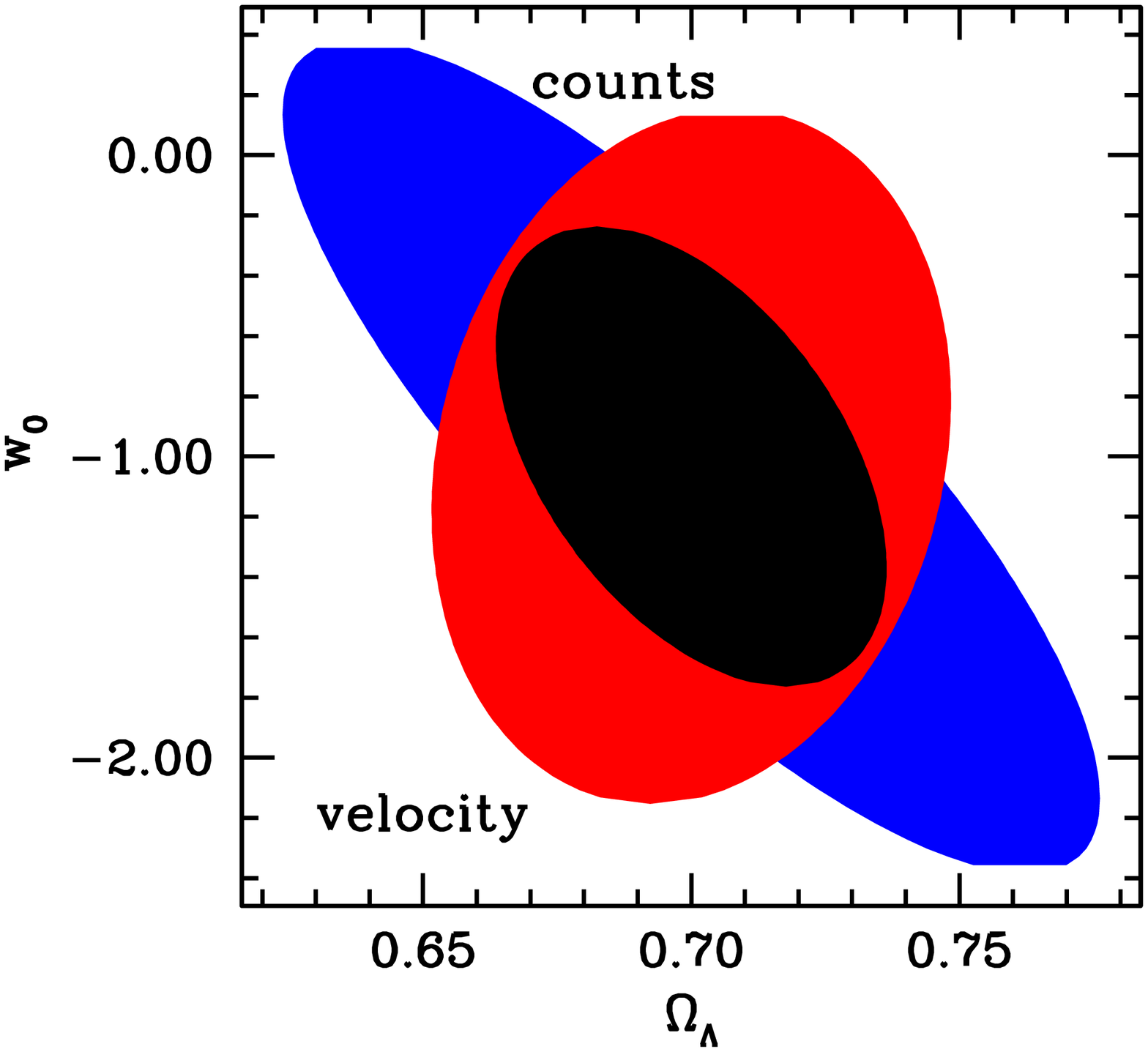}}
        \resizebox{85mm}{!}{\includegraphics{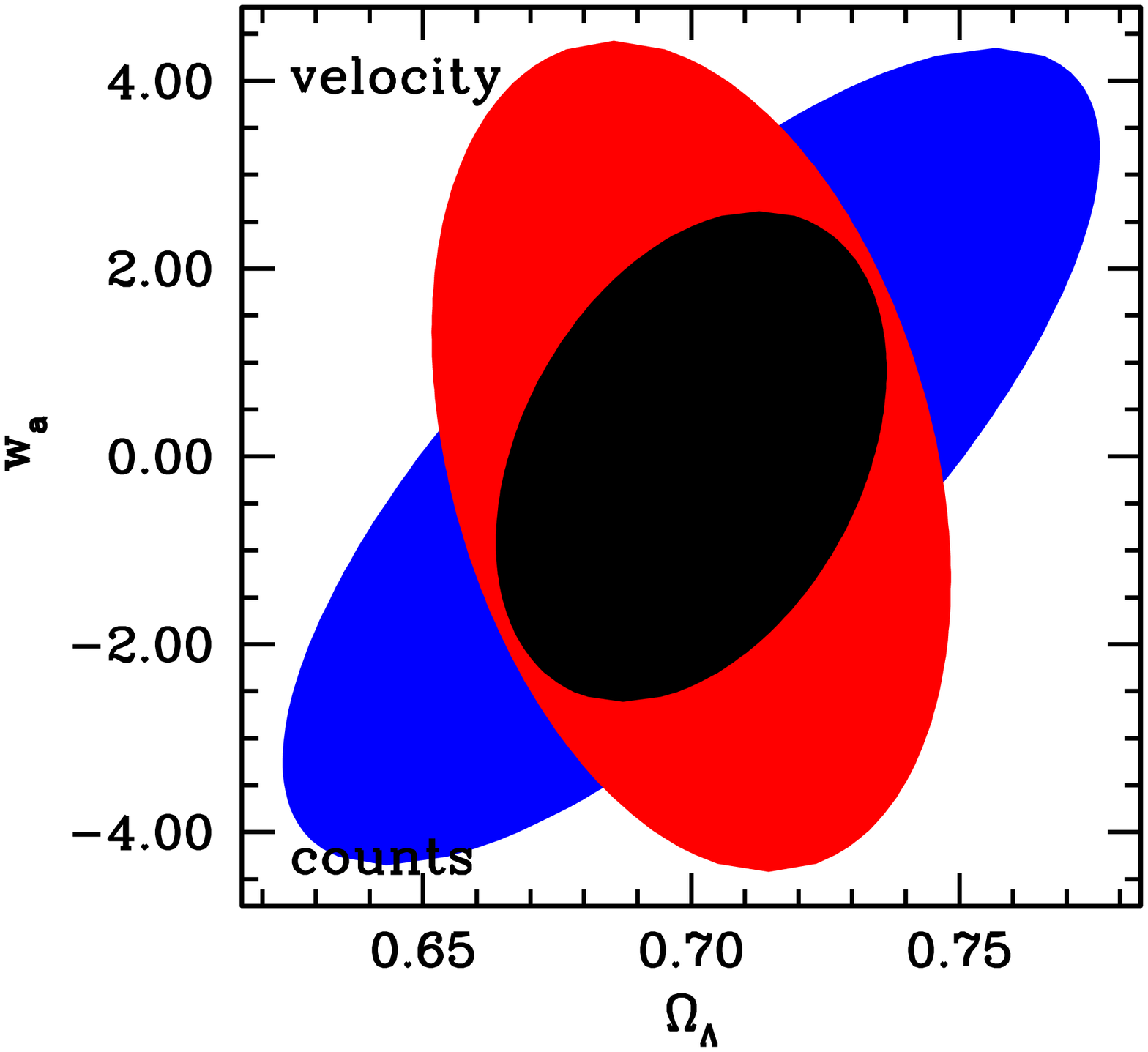}}\\
        \end{tabular}
    \caption{The relative complementarity of velocity and cluster counts. Shown are 1$\sigma$ error 
    ellipses in the $w_0-\Omega_{\Lambda}$ plane (left) and the $w_a-\Omega_{\Lambda}$ plane (right) 
    for 4000 clusters with normally-distributed velocity errors of $\sigma_v=1000 km/s$. The three ellipses 
    are for cluster velocities (red), cluster counts (blue) and the combination of both (black). Planck and  
    HST cosmological priors \citep{detf,hst} and a spatially flat cosmology are assumed.}
\label{complementarity}
  \end{center}
\end{figure}

For a given SZ survey, we can potentially obtain both cluster counts and cluster peculiar velocities. Given 
these two different data sources from the same survey, what is the joint constraint on dark energy 
parameters they provide? Consider a fiducial Stage II survey of 4000 galaxy clusters proposed by the 
DETF report
\citep{detf} (see Table~\ref{detf_surveys} for details), plus the addition of cluster velocities with
measurement error 
$\sigma_v=1000$ km/s, along with cosmic variance and Poisson errors to estimate
the mean pairwise peculiar velocity statistic $v_{ij}$. This is not a particularly stringent 
velocity error, and it is likely obtainable with currently
planned surveys with foreseeable follow-up observations or theoretical assumptions about cluster
properties.
Table~\ref{tab:v1000} gives the constraint on the dark energy parameters derived considering cluster
counts only, considering cluster velocities only, and the joint constraint from both. Also given are 
HST plus Planck prior constraints assuming a flat spatial geometry. We find cluster velocities provide a 
better constraint on $\Omega_{\Lambda}$ and $w_0$ than cluster counts, even for a measurement error 
of $\sigma_v=1000$ km/s. The constraint on $w_a$ is comparable for the two probes. The combined 
constraint is a factor of two better than the counts-only case for $\Omega_{\Lambda}$, $w_0$ and the 
Figure of Merit, and at
least a 60\% improvement for $w_a$. The relative complementarity between the two probes is shown in 
Figure~\ref{complementarity}.

We have assumed that the cluster velocity and cluster density observables are 
statistically uncorrelated. As they
will likely be obtained from the same set of clusters, it is reasonable to ask whether this is actually
true. A straightforward analytic calculation shows that the cross-correlation between 
velocity and density will be proportional to the
matter bispectrum, so we expect it to be small compared to the signal from the
velocity correlations, which are proportional to the matter power spectrum. We intend to
confirm this prediction from sets of large-volume numerical simulations when these
are available.

\begin{table*}
\caption{1$\sigma$ constraints for dark energy parameters for a fiducial cluster survey of 4000 clusters 
with velocity errors $\sigma_v=1000$ km/s, for cluster number counts, cluster velocities, and
the two combined. Planck and HST cosmological priors \citep{detf,hst} plus spatially flat cosmology 
assumed.}
\begin{center}
\begin{tabular}{ccccc}
\hline
Parameters & Priors & Counts & Velocity & Combined\\
\hline
$\Omega_{\Lambda}$[0.7] & 0.062 & 0.052 & 0.033 & 0.025\\
$w_0$[-1] & $-$ & 0.94 & 0.78 & 0.52 \\
$w_a$[0] & $-$ & 2.95 & 3.0 & 1.8 \\
FOM &  $-$ & 2.8 & 3.0 & 7.0 \\ 
\hline
\end{tabular}
\end{center}
\label{tab:v1000}
\end{table*}

\section{Comparison with DETF Proposed Experiments}

The Dark Energy Task Force report \citep{detf} considers four different potential probes to study dark
energy parameters: weak lensing(WL), baryon acoustic oscillations (BAO), cluster counts (CL) and SNIa
(SN) luminosity distance measurements. The relative merits of these probes have been discussed in 
detail in the DETF report both for ongoing and future projects. In this section we compare our fiducial 
velocity survey with each of the four DETF probes. To assess the advantage of adding cluster
mean pairwise peculiar velocity $v_{ij}$ as a dark energy probe, we have considered only the most optimistic forecasts for
the DETF surveys (i.e. survey assumptions that provide maximum constraint to the FOM assuming a flat 
universe  plus HST and Planck priors) for each Stage in the DETF report. Table~\ref{detf_surveys} gives a 
brief description of the DETF surveys considered here and our 
corresponding assumed cluster velocity surveys. We have used the actual Fisher
matrices used by the DETF team along with their priors for the following comparisons.


\begin{table*}
\caption{Parameters defining various surveys discussed in the DETF report \cite{detf}, plus various
cluster velocity surveys discussed here.}
\begin{center}
\begin{tabular}{cccccc}
\hline
Stages&VEL&WL&SNIa&Cl&BAO\\ \\\hline 
II & $N_{cl}=4000$, $f_{\rm sky}= 0.01$ & $f_{\rm sky}= 0.0042$ & SNLS & $N_{cl}=4000$ & None\\
 & $M_{min}> 2 \times 10^{14} M_{\odot}/h$ & & 700 SNIa & $f_{\rm sky}= 0.005$ & \\
 & z=0.1-1.4 & & z=0.1-1.0 & & \\ 
\\
\hline
\\
III & $N_{cl}=15000$ & DES & 2000 SNIa & $N_{cl}=30000$ & $f_{\rm sky}=0.1$\\
 &  $f_{\rm sky}= 0.05$ & $f_{\rm sky}= 0.1$ & Spectroscopy &  & \\ \\
\hline
\\
IV & $N_{cl}=30000$ & SKA-o & Space & $N_{cl}=30000$ & SKA-o\\
 & $f_{\rm sky}=0.1$ & $f_{\rm sky}= 0.5$ & 2000 SNIa & $f_{\rm sky}=0.5$ & $f_{\rm sky}= 0.5$\\
& &$z=0.1$--1.7 & & $z=0$--1.5 & \\ \\\hline
\end{tabular}
\end{center}
\label{detf_surveys}
\end{table*}

For the fiducial cluster velocity surveys, we have assumed that SZ surveys will be sensitive enough to 
detect the kSZ signal from all clusters with $M>2 \times 10^{14} M_{\odot}/h$. To be consistent with the 
DETF report, the total number of clusters for each survey corresponds to $\sigma_8= 0.9$. If 
$\sigma_8=0.76$ \citep{WMAP3} is used, then the corresponding number of clusters decreases by a
factor of 30\%. However,  a velocity survey is sensitive to only the number of detected clusters and not the 
volume of the survey. So our conclusions will still be valid if the survey area is increased to compensate 
for a lower value of $\sigma_8$. 

\begin{figure*}
  \begin{center}
    \begin{tabular}{cc}
       \resizebox{85mm}{!}{\includegraphics{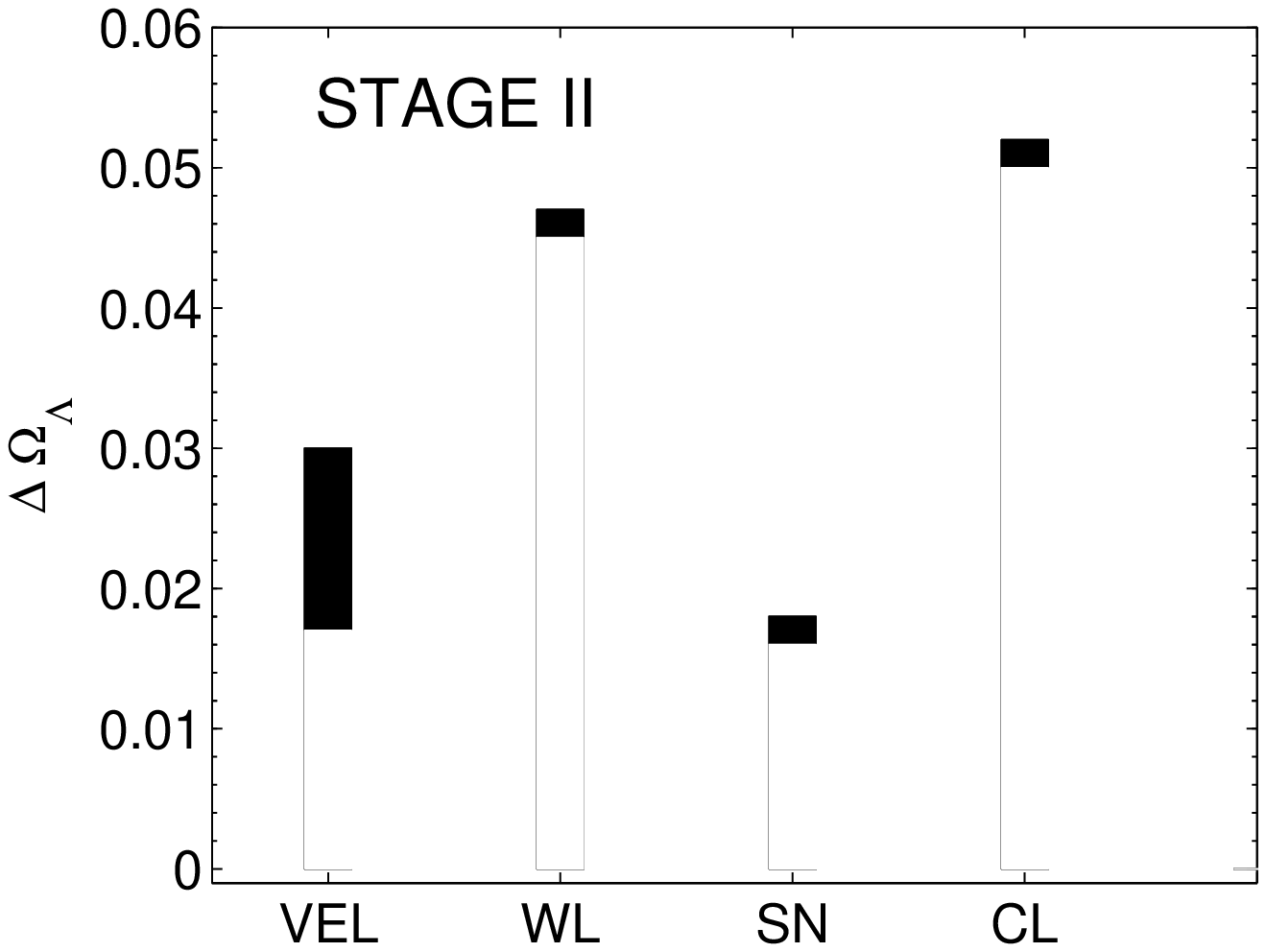}}
       \resizebox{85mm}{!}{\includegraphics{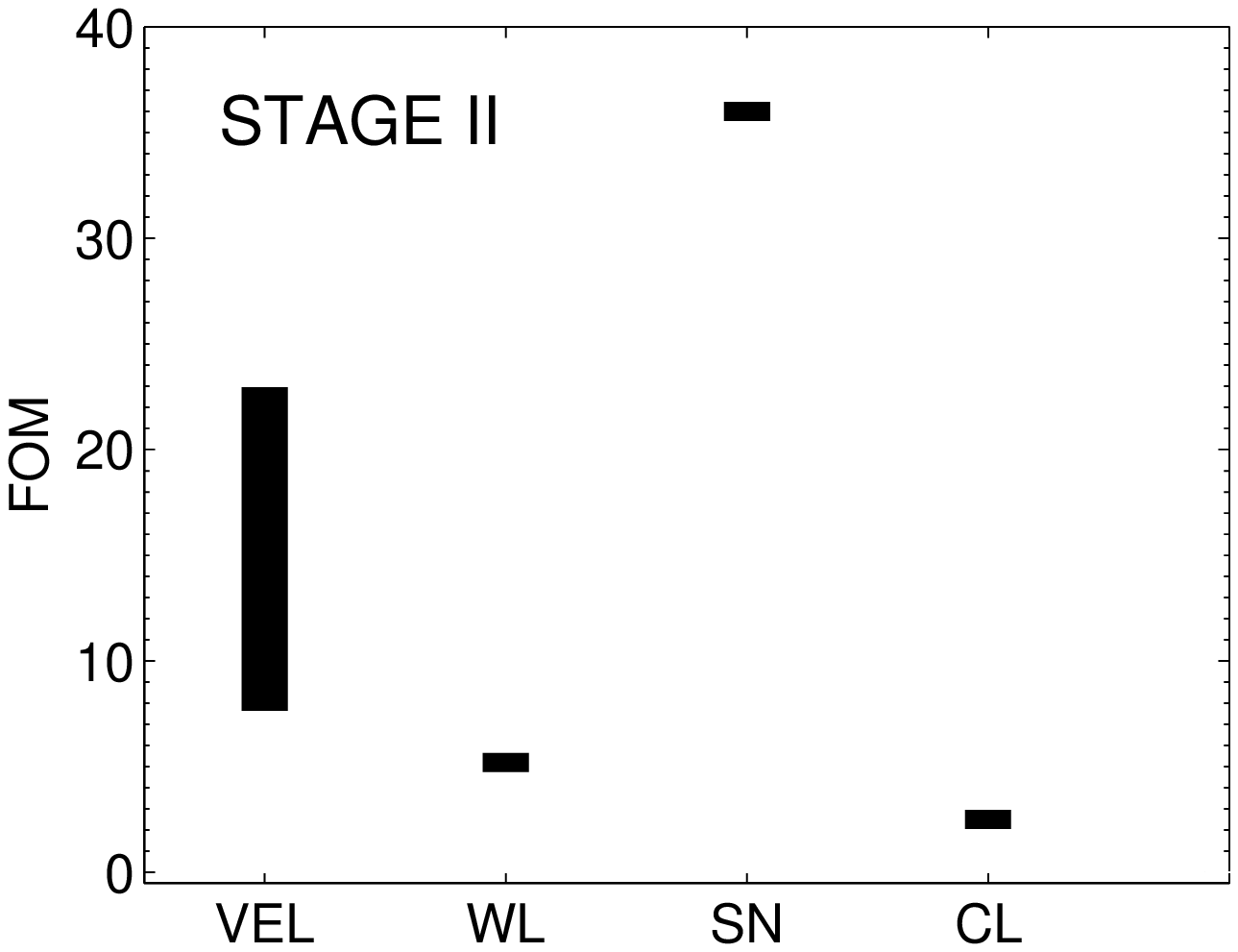}}\\
        \resizebox{85mm}{!}{\includegraphics{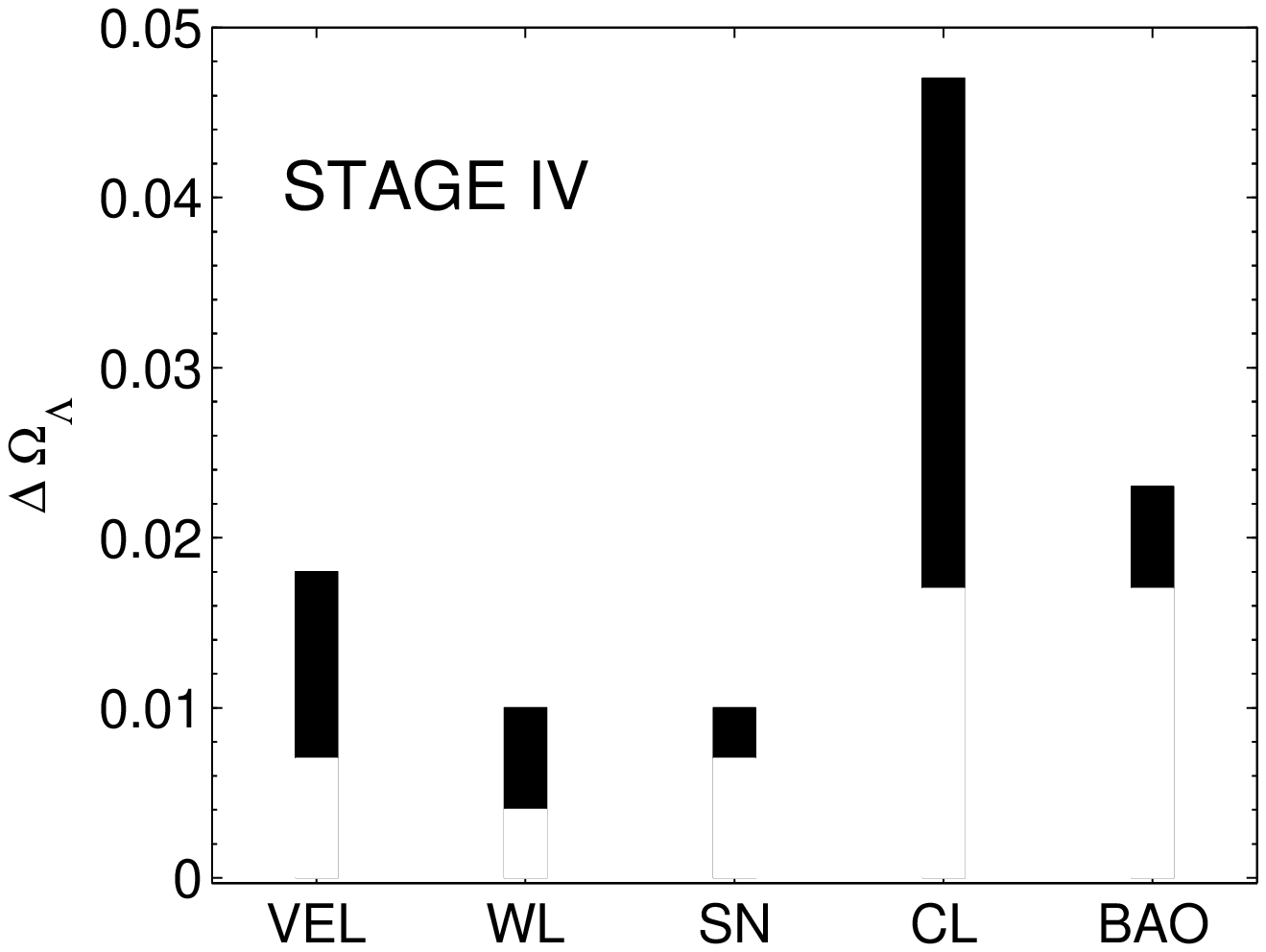}}
       \resizebox{85mm}{!}{\includegraphics{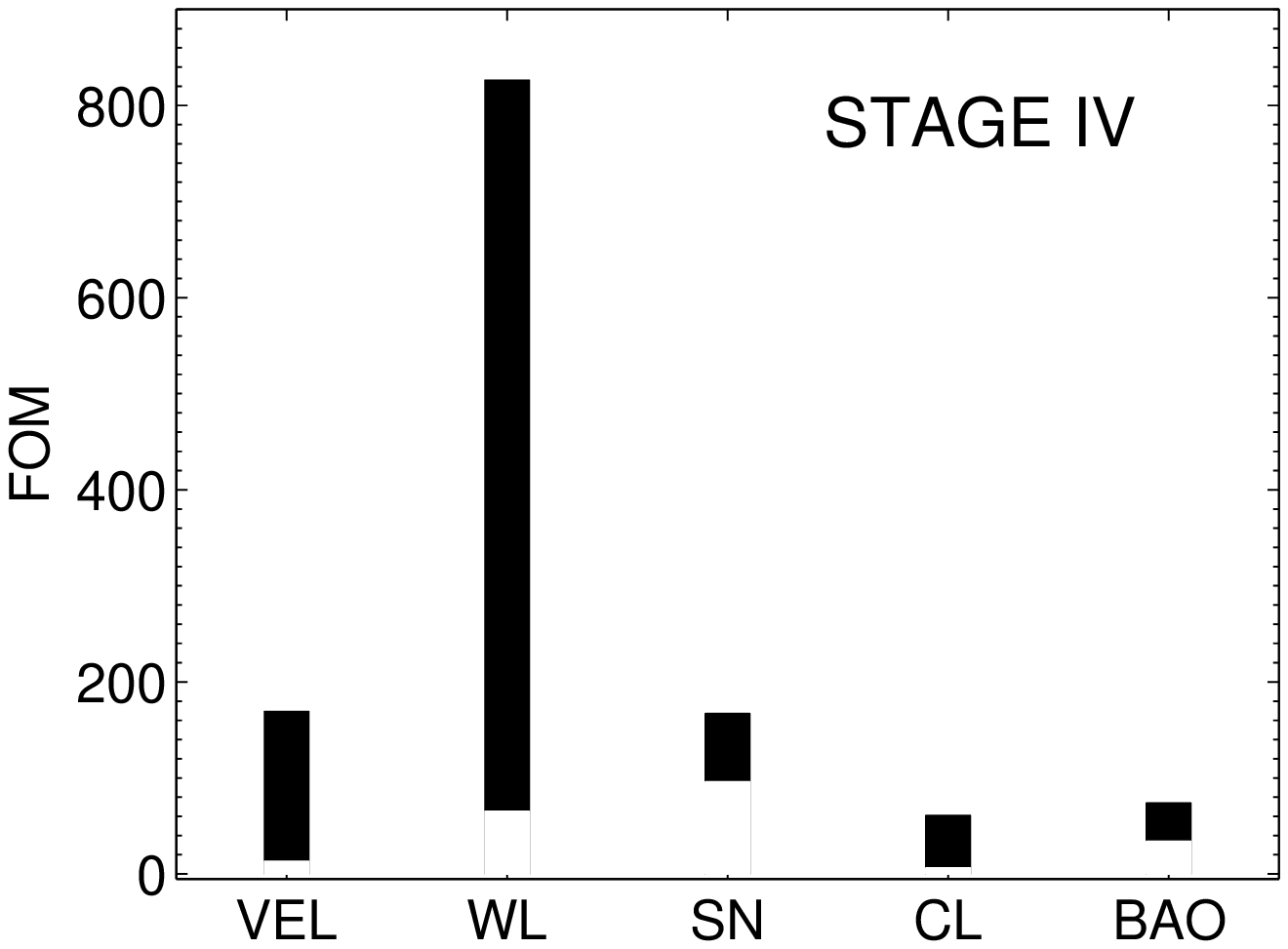}}\\
        \resizebox{85mm}{!}{\includegraphics{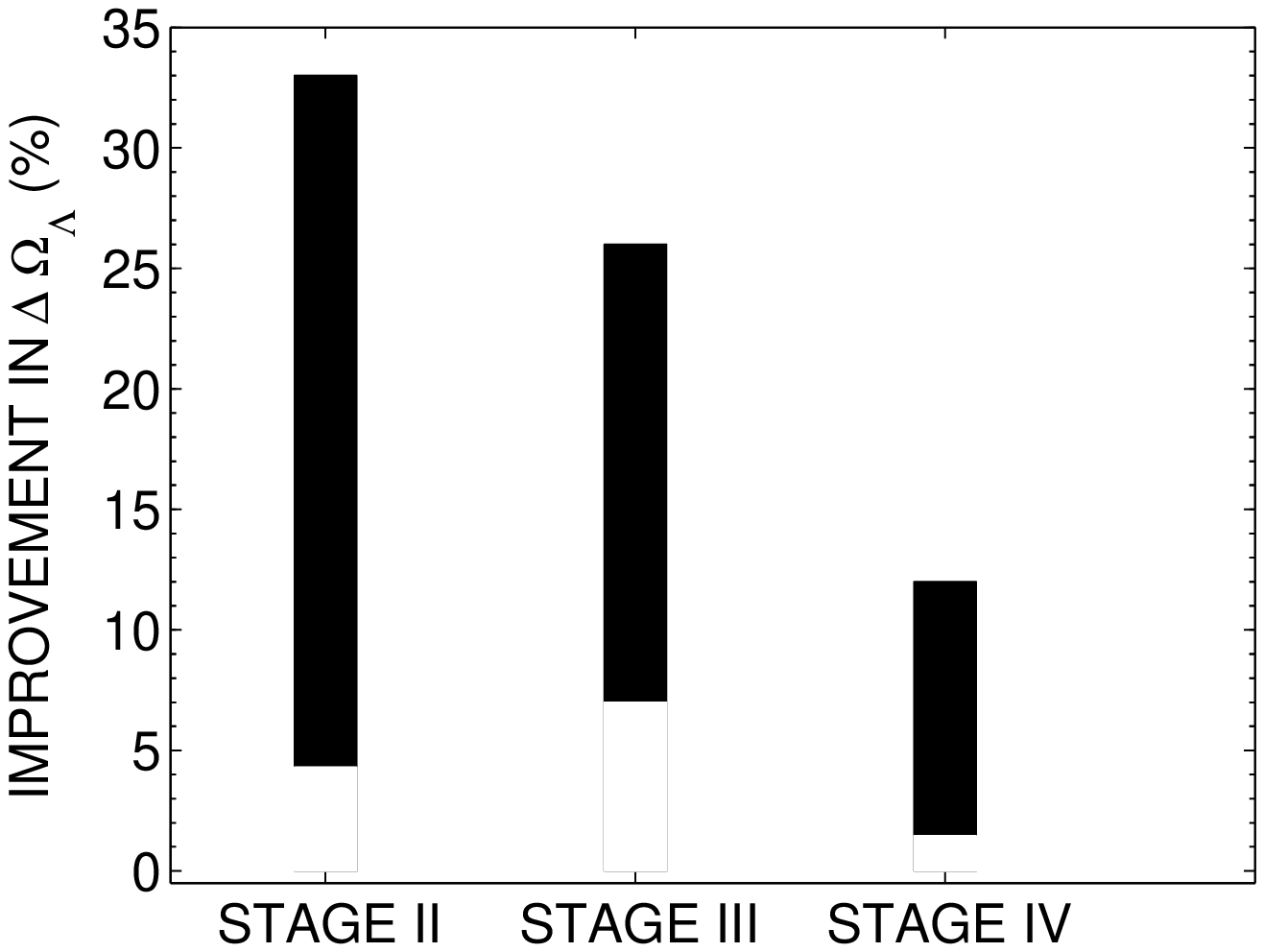}}
        \resizebox{85mm}{!}{\includegraphics{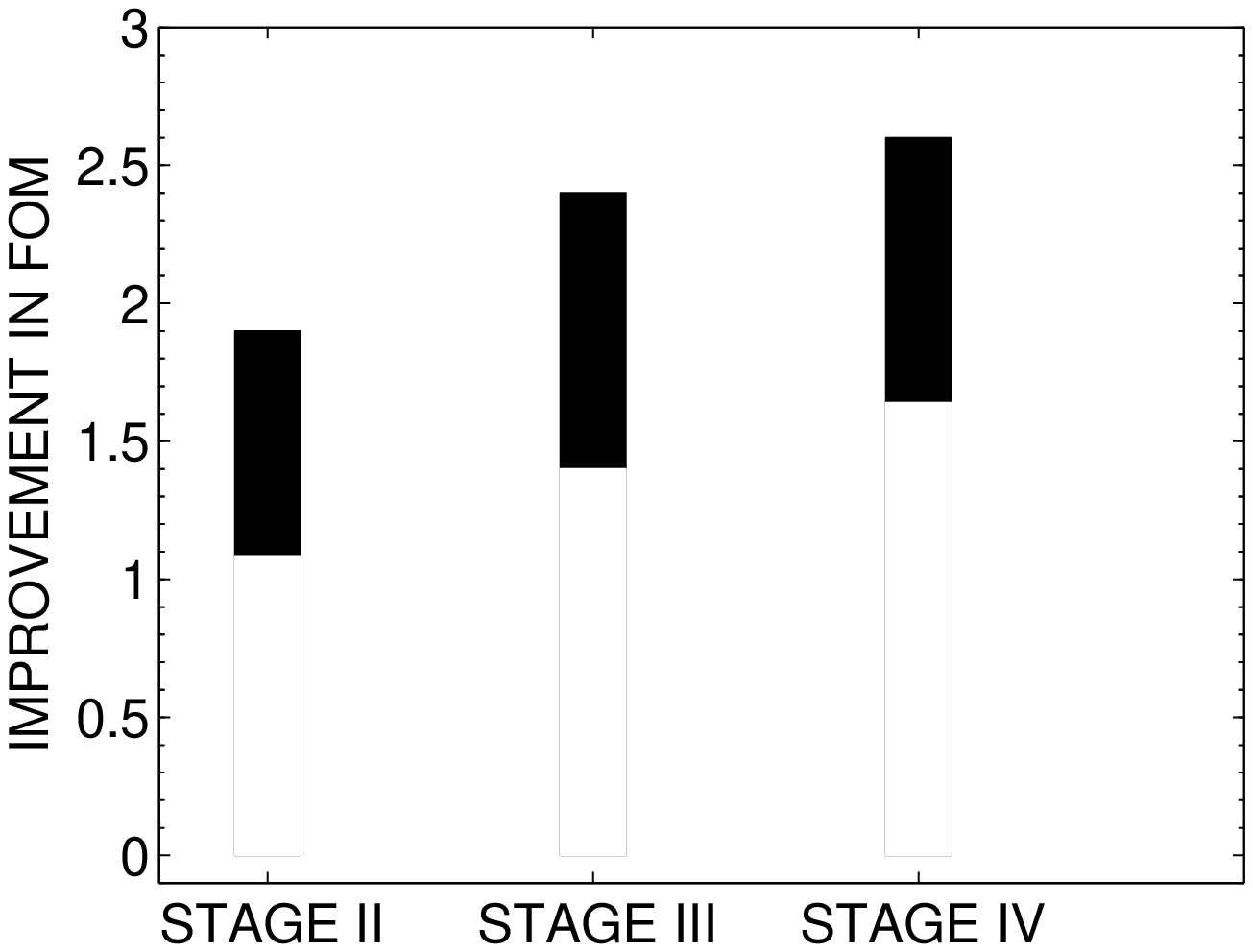}}\\        
        \end{tabular}
    \caption{A comparison of the error in the dark energy density $\delta \Omega_\Lambda$ and the dark
    energy figure of merit obtained from velocity statistics with that from DETF probes. The top two panels 
    are for Stage II experiments; the dark  region shows the range in the parameter error for the DETF-
    assumed ranges in the measurement errors. For cluster velocities we assume a range from
    $\sigma_v=200$ to 1000 km/sec. The middle panels show the results for Stage IV measurements. The 
    bottom panels show the relative improvement in parameter measurements at Stage IV when cluster 
    velocities are combined with all of the other DETF probes.}
         \label{fig:detf_compare}
  \end{center}
\end{figure*}

A comparison of velocity with other probes is shown in Figure~\ref{fig:detf_compare}. HST and Planck 
cosmological priors and spatially flat cosmology are assumed for all the probes. Each plot shows a range 
of parameter errors for each experiment, corresponding to cluster velocity measurement errors ranging
between 200 and 1000 km/sec, and other measurement errors as in the DETF report. At Stage II, velocity 
provides a competitive constraint on $\Omega_{\Lambda}$ compared to SNIa, and much better 
constraints than weak lensing or cluster number counts. Even a modest velocity survey would yield a 
factor of two better constrain on $\Omega_{\Lambda}$ than cluster counts or weak lensing. Cluster 
velocities also provide two to three times better constraints to the figure of merit  compared to weak 
lensing
or cluster counts at Stage II. Ultimately at Stage IV, however, weak lensing provides the most accurate 
measurements of dark energy density and the figure of merit. But constraints from velocity are competitive 
with those from supernovae and better than those from cluster counts or baryon acoustic oscillations. 
Stage II and III experiments yield an average 20\% improvement in cosmological parameter 
determination, and Stage IV about a 7\% improvement, when velocity information is combined with
the rest of the dark energy experiment results. This corresponds to an improvement by factors
of 1.5 to 2.5 in the dark energy figure of merit. These types of statistical comparisons of course
assume zero systematic errors; cluster velocities will ultimately be more valuable than these numbers
indicate, due to their completely different systematic errors from the other challenging techniques.
All of these methods will in the end be dominated by systematic, not statistical, errors.

\section{Discussion} 

The various studies of galaxy cluster peculiar velocities in this paper yield a number of interesting 
conclusions. The measurement of peculiar velocities of objects at cosmological distances
is of fundamental importance, as it directly probes the evolution of the gravitational
potential. The kinematic Sunyaev-Zeldovich effect in clusters of galaxies promises a
direct tracer of this signal, with errors largely independent of cluster redshift. 
Although the current uncertainty in velocity measurements is large with $\sigma_v \approx 1000$ km/s 
\citep{benson03} for individual clusters, upcoming multi-band experiments like ACT \citep{kosowsky03} or 
SPT \citep{ruhl05} with arcminute resolution and few $\mu$K sensitivity have the potential to measure 
peculiar velocities with velocity errors of a few hundred km/s for large samples of clusters, opening
a new window on the evolution of the universe. We have considered three separate cluster velocity 
statistics here, computing them using the halo model and comparing with numerical results. 
For surveys with thousands of cluster velocities with errors of a few hundred km/sec, 
dark energy constraints competitive
with other major techniques (cluster number counts, baryon acoustic oscillations, supernova redshift-
distance measurements, and weak lensing) can be obtained from the mean pairwise
peculiar velocity $v_{ij}$, with different systematic errors. Even
for velocity errors as large as 1000 km/s for individual clusters, a velocity catalog for  several thousand 
clusters can improve dark energy constraints from the corresponding cluster number counts by
a factor of two.

Throughout this work, we have simply assumed that cluster velocities can be extracted with
normal errors and no bias from
a Sunyaev-Zeldovich sky survey of sufficient angular resolution and low enough noise. Connecting the
measured SZ signal to the cluster velocity is a non-trivial task. The three ACT measured frequencies
at 145, 220, and 270 GHz have a degeneracy which prevents the cluster velocity from being determined
uniquely along with the cluster optical depth and temperature \cite{sehgal05,holder04}. This can be
remedied several ways, including adding other microwave bands \cite{holder04} or X-ray temperature 
measurements \cite{sehgal05}, or assuming cluster scaling relations between various measurable
quantities \cite{verde02,mccarthy03}. 

Further complications arise because the measured signal is not due only to the
Sunyaev-Zeldovich distortions, but also contains the blackbody primordial microwave fluctuations,
gravitational lensing of the microwave background, infrared and radio point sources which can
be correlated with galaxy cluster positions, and galactic dust (see \cite{sehgal07} for a description
of sky simulations incorporating all of these signals). The kinematic SZ signal must be separated
from all of the others via a combination of frequency and spatial filtering. With sufficient data, this
can clearly be done uniquely, but with limited wave bands, spatial resolution, and noise levels,
any kSZ signal extraction will be subject to some amount of measurement error. Evaluation of
this error for various observing strategies is important and we are currently pursuing it using
simulations. Even with perfect separation, internal cluster gas motions provide an
irredicible error floor for kSZ cluster velocity measurements of around 100 km/sec
\cite{nagai03,diaferio05}.

Component separation and other issues may also lead to systematic errors. We are
currently modeling systematic errors in velocity measurements in some detail, but it
is clear that at minimum, cosmological 
constraints based on cluster velocity measurements are much less
prone to systematic errors due to uncertainties in the relation between SZ distortion
and cluster mass than constraints
based on cluster number counts \cite{francis05,lima05}. This is potentially the dominant systematic
error for cluster number count studies, and largely mitigating it is one strong incentive
for pursuing cluster velocities as an alternative probe of dark energy. 
An additional advantage of using
cluster velocities is that the cluster velocity distribution function $n_v$ should be
symmetric with respect to positive and negative peculiar velocities, by homogeneity
of the universe. Departures from symmetry are easily diagnosed and can be used
as a monitor of unknown systematic errors. The downside of cluster velocities is that
the kSZ signal is much smaller than the thermal SZ signal, on the order of 5 to 10 $\mu$K
for large clusters with typical peculiar velocities. Separating this small signal from other
larger ones may lead to different systematic errors. But potential constraints on dark
energy from cluster velocities are good enough, and the other methods of measuring dark
energy properties are hard enough, that building a cluster velocity catalog with a different
set of systematic errors from other techniques is surely valuable. 

A number of further lines of work related to cluster velocities are worth pursuing. Here we have
considered three different galaxy cluster velocity statistics: the velocity probability distribution
function $n_v$, the mean pairwise velocity dispersion $v_{ij}$, and the velocity
correlation function $\langle v_iv_j\rangle$.  Each constrains well a different set of cosmological
quantities. We have not attempted a joint analysis, finding the combined cosmological constraints
from all three statistics: the correlations between the statistics are complicated, and no clear
way to derive them analytically presents itself. Proper joint constraints will require numerical
evaluation of the correlations between statistics from sets of large cosmological simulations,
which is feasible but demanding. A related question is the extent to which these three
statistics, which are convenient from a theoretical and observational point of view, exhaust
the useful cosmological information on dark energy constraints: are there other velocity
statistics which, when combined with these three using the correct correlations, would 
further tighten the constraints? This is an open, and challenging, question.

On the numerical front, we have performed limited tests comparing the VIRGO
simulation results with the halo-model expressions for the velocity statistics here,
finding reasonable agreement for the particular cosmological model the simulation
is based on. This is encouraging, but it would be reassuring to have explicit comparisons
between theory and simulation for a wider range of models. Such computations require
cosmological simulations over very large volumes, to capture a sufficient number of
clusters with large enough masses, but can be done with fairly low mass resolution,
since we only care about bulk cluster properties and not internal cluster details. Sets
of such simulations are currently in progress.

The kinematic SZ signal does not directly measure cluster peculiar velocity, but rather 
is proportional to a line-of-sight integral of the cluster gas' local peculiar velocity times its
local density. Thus the kSZ effect is actually proportional to the cluster gas momentum
with respect to the cosmic rest frame. We can sidestep the entire difficult observational issue of inferring 
cluster velocities from kSZ measurements by using cluster momenta instead. We then need theoretical 
calculations for the cluster momentum statistics corresponding to the velocity statistics considered here. 
Momentum statistics have the possibility of being just as cosmologically constraining, but easier to 
compare with observations. We have not found any suitable analytic approximations to the
cluster momentum statistics,  but this could also be evaluated numerically using large-volume, low-
resolution N-body simulations mentioned above. The other related issue is connecting the
cluster mass, which is used to evaluate cluster momenta in an N-body simulation, to the cluster gas
mass, which gives the SZ signal. We need to understand the extent to which the cluster gas fraction is 
constant, or the extent to which we can understand its statistical distribution. We have already
made initial steps to investigate this issue, finding, among other things, that
the gas fraction in galaxy groups is affected non-negligably by quasar feedback, which heats
the gas and suppresses star formation. However, at mass scales substantially below galaxy clusters,
the gas fraction appears to be reasonably independent of mass. Probing this relation for
clusters is a challenging computational issue, requiring sophisticated hydrodynamical
simulations  in much larger volumes to obtain information about galaxy clusters large enough to be of 
SZ interest. 

As with so many cosmological sources of information, the advent of the dark energy era has given a
new urgency to precision measurements. Galaxy cluster velocities, obtained via their kinematic
Sunyaev-Zeldovich signal, directly probe the growth of structure in the universe via gravitational
instability. The signals are small, but the advantages manifest. We firmly advocate that cluster
velocities should be added to the arsenal of tactics now trained on the dark energy issue.

\begin{acknowledgments}
We are grateful to Lloyd Knox and Wayne Hu for helpful discussions. We also thank Lloyd Knox, Jason 
Dick, and the Dark Energy Task Force for making available the DETF Fisher matrices. 
Andrew Zentner made useful suggestions related to complementarity between cluster counts and 
velocities, and Jeff Newman provided helpful background on cluster redshift measurements. 
This work has been supported by NSF grant AST-0408698 to the ACT project, and
by NSF grant AST-0546035.
\end{acknowledgments}

\onecolumngrid

\appendix

\section{Errors for the Probability Density Function}

\subsection{Poisson Error}

Let $N_z$ be the number of halos in redshift bin $z+\delta z$, and $N_v$ be the number of halos in both 
the redshift bin $z+\delta z$ and the velocity bin $v+\delta v$. In a given velocity bin,
the fractional density The observable in the normalized histogram of cluster velocities in a given redshift
bin is then $n_v=N_v/N_z$. Thus $n_v$ suffers from uncertainties in both numerator and denominator. 
We write the uncertainty in $n_v$ as 
\begin{equation}
\frac{\delta n_v}{n_v}= \frac{\delta N_v}{N_v}+\frac{\delta N_z}{N_z} \nonumber
\end{equation}
Assuming Poisson errors,  $\delta N_v= \sqrt{N_v}$ and $\delta N_z = \sqrt{N_z}$. We write
\begin{equation}
\delta n_v={\sqrt n_v}/{\sqrt N_z}+ n_v/{\sqrt N_z} =
\frac{{\sqrt n_v}[1+ {\sqrt n_v}]}{{\sqrt N_z}}
\end{equation}

\subsection{Cosmic Variance Error}

Write the cosmic covariance between two different bins $[v_i,z_i]$ and $[v_j,z_j]$ as $C^{n_v}_{ij}$; here 
$v_i$ denotes a particular velocity bin at an epoch of redshift $z_i$. $C^{n_v}_{ij}$ is defined as
\begin{equation}  
C^{n_v}(ij)= \left<(\hat{n}_{vi}- n_{vi})(\hat{n}_{vj}- n_{vj})\right>
\end{equation}
where $\hat{n}_v$ denotes the estimated PDF and $n_{vi}=n_v(v_i,z_i)$ etc. Using 
$n(m,\delta,{\bf x})= (1+b(m)\delta({\bf x}))\bar{n}(m)$ and 
$\hat{n}_v=V(r)^{-1}[\int d^3{\bf x} \int  dm\,mn(m|\bar{\delta},{\bf x})p(v\,|\,m,\delta)]/\int dm\,mn(m|\delta)$, 
\begin{equation}  
C^{n_v}(ij)= b(m_i,z_i)b(m_j,z_j)f_if_j\left<\delta({\bf x}_i,z_i) \delta^*({\bf x}_j,z_j)\right>
\end{equation}
where $\langle...\rangle$ denotes the ensemble average over the survey volume $V_\Omega$ and can 
be written as
\begin{equation}
\left<\delta_i \delta_j\right>= \frac{1}{V_\Omega}\int_{V_{\Omega}} d^3{\bf r}\int\int d^3{\bf x} d^3{\bf x'}
W({x})W({x'})\delta({\bf x},a)\delta({\bf x'},a')\delta^3_D({\bf x}-{\bf x'}-{\bf r})
\end{equation}
where
\begin{equation}
\delta({\bf x},a)\equiv D_a\delta({\bf x})= D_a\int d^3{\bf k}\, \delta({\bf k}) e^{i{\bf k}\cdot{\bf x}},
\end{equation}
$W(x)$ is the tophat window function defined after Eq.~(\ref{eq:massmoments}) and $\delta({\bf x})$ is
the field describing linear comoving density perturbations evolved to the present;
the three-dimensional Dirac delta distribution is written as $\delta^3_D({\bf x})$. 
We can then write
\begin{eqnarray}
\left<\delta_i \delta_j\right> &=& \frac{D_{a_i}D_{a_j}}{V_\Omega}
\int_{V_{\Omega}} d^3{\bf r}\int\int d^3{\bf x} d^3{\bf x'}W(x)W(x')
\delta({\bf x})\delta({\bf x'})\delta^3_D({\bf x}-{\bf x'}-{\bf r})\nonumber \\
&=&  \frac{D_{a_i}D_{a_j}}{V_\Omega}\int d^3{\bf r}\int\int d^3{\bf k} d^3{\bf k'}\delta(\bf k)\delta^*(\bf k')
e^{-i{\bf k}\cdot{\bf r}}h({\bf k}-{\bf k'},{\bf r}).
\end{eqnarray}
where we write conventionally \cite{power2d01}
\begin{equation}
h({\bf k},{\bf r})\equiv\frac{1}{V(r)}\int d^3{\bf x} W(x)W(|{\bf x}+{\bf r}|)e^{i{\bf k}\cdot{\bf x}}.
\label{hdef}
\end{equation}
In the limit of a survey region large compared to the scale $r$, 
$h({\bf k},{\bf r})\sim \delta^3_D({\bf k})$, $r\ll R_\Omega$ \citep{takada07,power2d01} with
the convenient notation $V_{\Omega} = 4 \pi R_{\Omega}^3/3$ for a spherical survey volume, giving
\begin{equation}
\int d^3{\bf x} W(x)W(|{\bf x}+{\bf r}|)e^{i({\bf k}-{\bf k'})\cdot{\bf x}}{\int d^3{\bf x} W^2(x)} 
\propto \delta^3_D({\bf k}-{\bf k'}).
\end{equation}
Then
\begin{equation}
\left<\delta_i \delta_j\right>= \frac{4\pi R_\Omega^2D_{ai}D_{aj}}{V_\Omega}
\int dk k^2P(k) j_1(kR_{\Omega}),
\end{equation}
so $C^{n_v}(ij)$ can be written as
\begin{equation}
C^{n_v}(ij)= \frac{3D_{a_i}D_{a_j}}{R_\Omega}n_i n_j \int dk k^2P(k)j_1(kR_{\Omega})
\end{equation}
where 
\begin{equation}
n_v(v,z)= \frac{\int dm mb(m,a){\bar n(m)}p(v|m,\delta,a)}{\int dm m {\bar n(m)}}
\end{equation}
which is equivalent to Eq.~(\ref{C_fij}).
The expression $p(v|m,\delta)$ is defined in Eq.~(\ref{eq:pvmd}).

\section{Errors for the Mean Pairwise Streaming Velocity}

\subsection{Poisson Error and Measurement Error}

We begin with Eq.~(\ref{v12est}) for the estimator of the mean pairwise streaming velocity.
Assume a particular velocity is measured with an accuracy $\delta v$. So the error 
$\delta v_{ij}$ in $v_{ij}$ can be written as
\begin{equation}
\frac{\delta v_{ij}}{v_{ij}}= \frac{\delta \Sigma_{ij}[v_i-v_j]}{ \Sigma_{ij}[v_i-v_j]}+ \frac{\delta n_p}{n_p},
\end{equation}
so that
\begin{equation}
\delta v_{ij}= \frac{\sqrt{2}\left[\Sigma_i \delta v^2_i\right]^{1/2}}{n_p}
+ \frac{v_{ij}}{\sqrt n_p}
= \frac{1}{\sqrt n_p}\left(\sqrt{2} \sigma_v + v_{ij}\right)
\end{equation}
where we have used $\delta n_p= \sqrt n_p$ assuming a Poisson distribution, and
\begin{equation}
\delta  \Sigma_{ij}\left[v_i-v_j\right]= 
 \sqrt{2}[\delta v^2_1+ \delta v^2_2+ ...+ \delta v^2_{n_p}]^{1/2} = \sqrt{2}{\sqrt n_p} \sigma_v.
\end{equation}
Here the individual velocity errors are added in quadrature and the last line follows from the central limit 
theorem.

\subsection{Cosmic Variance Error}

The cosmic covariance for mean pairwise streaming velocity between two separation and redshift
bins $[r_p, z_p]$ and $[r_q, z_q]$ can be written as
\begin{eqnarray}
C^{v_{ij}}(pq)&=& \left<\left(v_{ij}(p)-{\hat v_{ij}(p)}\right)\left(v_{ij}(q)-{\hat v_{ij}(q)}\right)\right> 
=  \left<{\hat v_{ij}}(p){\hat v_{ij}}(q)\right>- v_{ij}(p)v_{ij}(q)
\end{eqnarray}
where ${\hat v_{ij}}$ is the estimated mean pairwise streaming velocity from the survey volume and 
$v_{ij}$ is its cosmic mean value, $\langle{\hat v_{ij}}\rangle= v_{ij}$.
Using the expression for mean pairwise streaming velocity given in Eq.~(\ref{v12}), the above expression 
can be written as
\begin{eqnarray}
C^{v_{ij}}(pq)&=&\frac{1}{1+\xi^{\rm halo}(r_p,a_p)}
\left[\frac{2}{3}r_pH(a_p)a_p\left(\frac{d \ln D_a}{d \ln a}\right)_{a_p}\right]
\frac{1}{1+\xi^{\rm halo}(r_q,a_q)}
\left[\frac{2}{3}r_qH(a_q)a_q\left(\frac{d \ln D_a}{d \ln a}\right)_{a_q}\right]\nonumber\\
&&\qquad\qquad\qquad\qquad
\times\left[\left<{\hat{\bar\xi}}(r_p){\hat{\bar\xi}}(r_q)\right>- {\bar\xi}(r_p){\bar\xi}(r_q)\right],
\label{C_vij_pq1}
\end{eqnarray}
where $\hat{\bar\xi}$ is an estimator for the volume-averaged correlation function 
\begin{equation}
\bar\xi(r) \equiv \frac{1}{V(r)}\int_0^r dr'\, r'^2\xi(r').
\end{equation}
An estimator $\hat\xi(r)$ for the two-point correlation function $\xi(r)$ is
\begin{equation}
\hat\xi(r) = \frac{1}{V(r)}\int d^3{\bf x'} W(x')\int d^3{\bf x} W(x) \delta({\bf x})\delta({\bf x'})
\delta^3_D({\bf x} - {\bf x'} - {\bf r}),
\label{xi_def}
\end{equation}
so an estimator for the volume-averaged correlation function can be written as
\begin{equation}
{\hat{\bar\xi}}(r)= \frac{1}{V(r)}\int_{V(r)} d^3{\bf r'}\frac{1}{V((r')}\int d^3{\bf x} W(x)
\int d^3{\bf x'} W(x')\delta({\bf x})\delta({\bf x'})\delta^3_D({\bf x}-{\bf x'}-{\bf r'})
\end{equation}
where the survey volume is given by $V(r)\equiv\int d^3{\bf x}\,W(x)W(|{\bf x}+{\bf r}|)$ for
a normalized window function $\int d^3{\bf x}\, W(x)=1$. 
Fourier transforming $\delta(x)$, we can write
\begin{eqnarray}
{\hat {\bar \xi}}(r)&=&  \frac{1}{V(r)}\int_{V(r)} d^3{\bf r'}\frac{1}{V(r')}\int d^3{\bf r'}\int d^3{\bf x} W(x)
\int d^3{\bf x'} W(x')\delta^3_D({\bf x}-{\bf x'}-{\bf r'})
\int\int d^3{\bf k}d^3{\bf k'} \delta({\bf k})\delta^*({\bf k'}) 
e^{i({\bf k}\cdot {\bf x}-{\bf k'}\cdot{\bf x'})}\nonumber\\
&=& \frac{1}{V(r)} \int_0^r d^3{\bf r'} \int\int d^3{\bf k} d^3{\bf k'}\delta({\bf k})\delta^*({\bf k'})
e^{-i{\bf k}\cdot{\bf r'}}h({\bf k}-{\bf k'},{\bf r'}).
\label{xi_est}
\end{eqnarray}
Using $\langle\hat{\bar \xi}(r)\rangle={\bar \xi}(r)$, we can then write 
\begin{eqnarray}
C^{\bar \xi}(pq)&=& \left[\left<{\hat{\bar \xi}}(r_p){\hat{\bar \xi}}(r_q)\right>- {\bar \xi}(r_p){\bar \xi}(r_q)\right] 
\nonumber\\
&=&  \frac{1}{V(r_p)V(r_q)}\int_0^{r_p}  d^3{\bf r} e^{-i{\bf k}\cdot{\bf r}}h({\bf k}-{\bf k'},{\bf r})
\int_0^{r_q}  d^3{\bf r'} e^{-i{\bf k}\cdot{\bf r'}}h^*({\bf k}-{\bf k'},{\bf r'})\nonumber\\
&\times& \int d^3{\bf k} \int d^3{\bf k'}\int d^3{\bf k_1} \int d^3{\bf k'_1}\left[\left<\delta({\bf k})\delta^*({\bf k'})
\delta({\bf k_1})\delta^*({\bf k'_1})\right>- \left<\delta({\bf k})\delta^*({\bf k'})\right>
\left<\delta({\bf k_1})\delta^*({\bf k'_1})\right>\right].
\label{C_xi_pq1}
\end{eqnarray}
The term in brackets can be written as 
\begin{equation}
[...]= \delta^3_D({\bf k}+{\bf k_1})P(k)\delta^3_D({\bf k'}+{\bf k'_1})P(k')+ \delta^3_D({\bf k}-{\bf k'_1})P(k)
\delta^3_D({\bf k'}-{\bf k_1})P(k').
\end{equation}
Substituting this expression into Eq.~(\ref{C_xi_pq1}) gives
\begin{equation}
C^{\bar \xi}(pq)=  \frac{1}{V(r_p)V(r_q)}\int d^3{\bf k} \int d^3{\bf k'} P(k)P(k')\left(e^{i{\bf k}\cdot({\bf r}-{\bf r'})}
+ e^{-i{\bf k}\cdot{\bf r}-i{\bf k'}\cdot{\bf r'}}\right)
\int_0^{r_p} d^3{\bf r} h({\bf k}-{\bf k'},{\bf r})\int_0^{r_q} d^3{\bf r'} h^*({\bf k}-{\bf k'},{\bf r'}). 
\label{C_xi_pq2}
\end{equation}
As in the previous appendix, for large surveys such that $r << R_\Omega = (3V_\Omega/4\pi)^{1/3}$,
$h({\bf k}-{\bf k'},{\bf r})\sim \delta^3_D({\bf k}-{\bf k'})$ and  \citep{takada07,power2d01} 
\begin{equation}
hh^*= \frac{\int d^3{\bf x} W^2(x)W(|{\bf x}+{\bf r}|)W(|{\bf x}+{\bf r'}|)}{V(r_p)V(r_q)}
\sim \frac{1}{V_{\Omega}}.
\end{equation}
So Eq.~(\ref{C_xi_pq2}) can be written as
\begin{eqnarray}
C^{\bar \xi}(pq)&=& \frac{1}{V_\Omega V(r_p)V(r_q)} \int d^3{\bf k} |P(k)|^2\int_0^{r_p}
\int_0^{r_q}d^3{\bf r} d^3{\bf r'}\left(e^{i{\bf k}\cdot({\bf r}-{\bf r'}}
+ e^{-i{\bf k}\cdot({\bf r}+{\bf r'})}\right)\nonumber\\
&=& \frac{8\pi}{V_\Omega r_p r_q}\int dk k^2 |P(k)|^2j_1(kr_p)j_1(kr_q)
\label{C_xi_pq3}
\end{eqnarray}
Substituting the above result in Eq~(\ref{C_vij_pq1}), we obtain the final expression for cosmic covariance 
as
\begin{equation}
C^{v_{ij}}(pq)=\frac{32\pi}{9V_\Omega}\frac{H(a_p)a_p}{1+\xi^{\rm halo}(r_p,a_p)}
\frac{H(a_q)a_q}{1+\xi^{\rm halo}(r_q,a_q)}\left(\frac{d \ln D_a}{d \ln a}\right)_{a_p}
\left(\frac{d \ln D_a}{d \ln a}\right)_{a_q}\int dk k^2 |P(k)|^2j_1(kr_p)j_1(kr_q).
\label{C_vij_pq2}
\end{equation}
On scales of interest, $\xi^{\rm halo} \ll 1$, so Eq.~(\ref{C_vij_pq2}) reduces to Eq.~(\ref{C_vij_cosmic}).

\section{Errors for the Velocity Correlation Function}

\subsection{Poisson Error and Measurement Error}

The expression for the perpendicular velocity correlation $\langle v_iv_j \rangle_\perp(r)$ for a particular 
separation $r$ can be written as 
\begin{equation}
\langle v_iv_j \rangle(r)=\frac{\Sigma_{ij}[v_iv_j]_\perp}{n_p}
\end{equation}
where we abbreviate 
$[v_iv_j]_\perp\equiv([{\bf r_i}-{\bf r_j}]\times {\bf v}_i)\cdot([{\bf r_i}-{\bf r_j}]\times{\bf v}_j)$
the product of the velocity components
perpendicular to the direction connecting the two positions.
As before, ${\bf v}_i$ is the velocity of halo $i$, which is measured with a normal error in its
magnitude of $\delta v$, and $n_p$ is the number of pairs in the survey volume for a given separation 
distance $r$. For the rest of the appendix, we drop the perpendicular subscript for convenience.
So the measurement error in $\langle v_iv_j \rangle$ can be written as
\begin{eqnarray}
\langle v_iv_j \rangle+ \delta \langle v_iv_j \rangle&=& \frac{1}{n_p}
\Sigma_{ij}[v_iv_j+2v_j \delta v_i+\delta v_i \delta v_j] \nonumber \\
\delta \langle v_iv_j \rangle&=& \frac{1}{n_p}\Sigma_{ij}[2v_j\delta v_i+\delta v_i \delta v_j]\nonumber \\
&=& \frac{1}{n_p} \Sigma [\delta (v^2)+ (\delta v)^2]
\end{eqnarray}
Similarly, the Poisson error is $\langle v_iv_j \rangle[\delta n_p/n_p]= \langle v_iv_j\rangle/{\sqrt n_p}$.

\subsection{Cosmic Variance Error}

The cosmic covariance for the velocity correlation function between two bins $[r_p, z_p]$ and $[r_q, z_q]$, 
one of separation $r_p$ at epoch $z_p$ and the other of separation $r_q$ at redshift $z_q$, can be 
written as
\begin{eqnarray}
C^{\langle v_iv_j \rangle}(pq)&=& \left<\left(\left<v_iv_j\right>(p)
-\left<{\langle\widehat{v_iv_j}\rangle}\right>(p)\right)\left(\left<v_iv_j\right>(q)
-\left<{\widehat{v_iv_j}}\right>(q)\right)\right> \nonumber \\
&=& \left<\widehat{v_iv_j}\right>(p)\left<\widehat{v_iv_j}\right>(q) - \left<v_iv_j\right>(p)\left<v_iv_j\right>(q)
\end{eqnarray}

As in the case of $v_{ij}(r)$, we first derive an estimator for $v_iv_j(r)$. In linear theory, 
$v(k) = \delta(k)/k$, so $v(x) = \int d^3{\bf k}[\delta(k)/k]\exp(i{\bf k}\cdot{\bf x})$. Then an estimator
$\widehat{v_iv_j}(r)$ measured at a  separation ${\bf r}$ is
\begin{eqnarray}
\widehat{v_iv_j}(r) &=& \frac{1}{V(r)}\int d^3{\bf x'} W(x')v{\bf x'}\int d^3{\bf x} W(x)v({\bf x'})
\delta^3_D({\bf x}-{\bf x'} - {\bf r})\nonumber\\
&=& \int\int d^3{\bf k}d^3{\bf k'}\frac{\delta({\bf k})\delta^*({\bf k'})}{kk'} e^{-i{\bf k}\cdot{\bf r'}} 
h({\bf k}-{\bf k'}).
\label{vij_est_lin1}
\end{eqnarray}
The only difference between Eq.~(\ref{vij_est_lin1}) and Eq.~(\ref{xi_est}) is the added
factor of $kk'$ in the denominator. 

The expression for the velocity correlation function given in Eq.~(\ref{v1v2perp}) consists of two terms, 
expressions for which are given in Eqs.~(\ref{I1}) and (\ref{I2}). For simplicity, here we derive the cosmic 
covariance of the first term using the linear theory expression for the velocity correlation, 
Eq.~(\ref{vij_est_lin1}); the derivation can be easily extended to the halo model expression for 
$\langle v_iv_j \rangle$ given in Eq.~(\ref{v1v2perp}). As argued before, the second term in 
Eq.~(\ref{v1v2perp}) can be neglected compared to the first term because $\xi(r)$ is negligible at 
separations of interest for $r>30$ Mpc.  The linear theory counterpart for Eq.~(\ref{v1v2perp}) can be 
written as 
\begin{equation}
\langle{\hat T_1}\rangle(r,a)= \left[H(a)\frac{d \ln D_a}{d \ln a}aD_a\right]^2\frac{1}{3V(r)} \int_0^r d^3{\bf r'} 
\int\int d^3{\bf k} d^3{\bf k'}\frac{\delta({\bf k})\delta^*({\bf k'})}{kk'}e^{-i{\bf k}\cdot{\bf r'}}
h({\bf k-k'},{\bf r'}).
\label{T1_est_lin}
\end{equation}
Note that this integrand is similar to that in to Eq.~(\ref{v1v2perp}), apart from the  
halo number density and bias factors. The factor of $1/3$ in Eq.~(\ref{T1_est_lin}), compared
to Eq.~(\ref{xi_est}), is because only the radial velocity components are considered. 
Proceeding analogously to Eqs.~(\ref{xi_est}) to (\ref{C_xi_pq2}), we obtain
\begin{equation}
C^{T_1}(pq) =  a_p^2a_q^2 D^2_{a_p}D^2_{a_q} H^2(a_p)H^2(a_q)
\left[\frac{d \ln D_a}{d \ln a}\right]^2_{a_p}\left[\frac{d \ln D_a}{d \ln a}\right]^2_{a_q}
\frac{64\pi^2}{V_\Omega^2}\int dk  P(k)^2\frac{j_1(kr_p)}{kr_p}\frac{j_1(kr_q)}{kr_q}
\end{equation}
This is the  cosmic covariance for the linear theory counterpart of Eq.~(\ref{v1v2perp}). 
Including the extra halo model factors gives Eq.~(\ref{C_vivj_cosmic}).

\bibliography{paper3v3PRD}

\end{document}